\documentclass{article}

\usepackage{arxiv}

\usepackage[utf8]{inputenc} % allow utf-8 input
\usepackage[T1]{fontenc}    % use 8-bit T1 fonts
\usepackage{hyperref}       % hyperlinks
\usepackage{url}            % simple URL typesetting
\usepackage{booktabs}       % professional-quality tables
\usepackage{amsfonts}       % blackboard math symbols
\usepackage{nicefrac}       % compact symbols for 1/2, etc.
\usepackage{microtype}      % microtypography
\usepackage{lipsum}

\usepackage{float}
\usepackage{hanging} %hangpara
\usepackage{xcolor,colortbl} 
\usepackage{array}
\usepackage{tikz, tikz-qtree}
\usetikzlibrary{shapes.geometric, arrows, positioning, chains, trees}
\usepackage{url}
\def\UrlBreaks{\do\/\do-}
\expandafter\def\expandafter\UrlBreaks\expandafter{\UrlBreaks%  save the current one
  \do\/\do-\do.\do+}

\usepackage{makecell}
\usepackage{enumitem,amssymb}
\newlist{todolist}{itemize}{1}
\setlist[todolist]{label=$\square$,leftmargin = 1cm}
\usepackage[shortcuts]{extdash}

\usepackage[mode=buildnew]{standalone}% requires -shell-escape
\usepackage{tikz}
\usepackage{caption}
\usepackage{chngpage} % allows for temporary adjustment of side margins

\usepackage[all]{nowidow}

\usepackage{adjustbox}
\usepackage{longtable, lscape}

\title{Checklist for responsible deep learning modeling of~medical images based on~COVID-19 detection studies}

\author{
  Weronika Hryniewska \\
  Faculty of Mathematics and Information Science \\
  Warsaw University of~Technology\\
  \texttt{w.hryniewska@mini.pw.edu.pl} \\
   \And
  Przemysław Bombiński \\
  Department of~Pediatric Radiology\\
  Medical University of~Warsaw\\
  \texttt{przemyslaw.bombinski@uckwum.pl} \\
   \And
   Patryk Szatkowski \\
   Department of~Pediatric Radiology\\
   Medical University of~Warsaw\\
   \texttt{szpatryk@poczta.onet.pl} \\
   \And
   Paulina Tomaszewska \\
   Faculty of Mathematics and Information Science \\
   Warsaw University of~Technology\\
   \texttt{paulina.tomaszewska3.dokt@pw.edu.pl} \\
   \And
   Artur Przelaskowski \\
   Faculty of Mathematics and Information Science \\
   Warsaw University of~Technology\\
   \texttt{a.przelaskowski@mini.pw.edu.pl } \\
   \And
   Przemysław Biecek \\
   Faculty of Mathematics, Informatics and Mechanics \\
   University of Warsaw \\
   Faculty of Mathematics and Information Science \\
   Warsaw University of Technology \\
   \texttt{przemyslaw.biecek@pw.edu.pl} \\
}

\begin{document}
\maketitle

\begin{abstract}
The sudden outbreak and~uncontrolled spread of~COVID\=/19 disease is one of~the~most important global problems today.
In~a~short period of~time, it has led to the~development of~many deep neural network models for~COVID\=/19 detection with modules for~explainability.
In~this work, we carry out a~systematic analysis of~various aspects of~proposed models.
Our analysis revealed numerous mistakes made at different stages of~data acquisition, model development, and~explanation construction.
In~this work, we overview the~approaches proposed in~the~surveyed Machine Learning articles and~indicate typical errors emerging from the~lack of~deep understanding of~the~radiography domain. We present the~perspective of~both: experts in~the~field - radiologists and~deep learning engineers dealing with model explanations.
The final result is a~proposed checklist with the minimum conditions to be met by~a~reliable COVID\=/19 diagnostic model.
\end{abstract}

% keywords can be removed

\section{Introduction}
\label{sec:introduction}

COVID\=/19 is a~fast spreading disease of~highly contagious nature caused by~the~SARS-CoV-2 virus from the~coronavirus group. At the~end of~January 2020, the~World Health Organization (WHO) declared a~global health emergency and~one and~a~half months later, a~pandemic. By September 25, 2020, 32,110,656 confirmed cases and~980,031 deaths had been documented. From~a~public health perspective, due to the~lack of~proper medicines, early detection of~COVID\=/19 and~patient isolation are crucial. Hospitals are crowded with~the~exponentially growing number of~patients as available resources are limited.

Currently, reverse transcription polymerase chain reaction (RT\=/PCR) is the~gold standard used to diagnose COVID\=/19 infection \cite{Xie}. However, the~results of~RT\=/PCR can be affected by~sampling errors and~low viral load \cite{Xie}. As~a~result, these tests suffer from~high rates of~false negatives (with sensitivity of~71\% \cite{Fang_Yicheng} or~69\% \cite{Wong_Ho}) and~may need to be conducted two or~more times before the results are finally confirmed \cite{Corman}.

In~many articles, chest imaging is considered a~suitable tool for~early COVID\=/19 screening \cite{Li2020, Kong2020}. The~point is~that sensitivity of~computed tomography (CT) scan tests can reach 98\%, which is much higher than RT\=/PCR tests \cite{Fang2020}. Moreover, due to the~fact that on~CT images ground-glass opacities are visible earlier than pulmonary consolidation \cite{Chung2020}, radiologists can assess the~stage of~COVID\=/19. Unfortunately, CT scanners are not widely available. Study \cite{Bai2020a}, indicate that distinguishing between COVID\=/19 and~viral pneumonia is a~challenging task. However, it is worth noticing that the overall process of~undertaking chest imaging to getting the first results is much shorter than in case of~RT\=/PCR. The~screening takes approximately 15 seconds \cite{Wong2020} (in terms of~X\=/ray) to 21.5 \cite{Huang2020battle} minutes on~average (for~CT) to complete. Taking the sample for~RT\=/PCR test is a fast procedure. The difference is not the~time of~undertaking the~test/screening but the~time needed to get the~first results. In case of~RT\=/PCR, it may take from~several hours up~to several days \cite{Long2020} as nucleic acid amplification must happen before the~start of~samples' analysis. The issue is important as the~patient has to be isolated until receiving the~test result.

In addition, X\=/rays are cheaper, more available worldwide, and~less harmful than CTs because the~radiation dose is smaller. Due to the~existence of~portable devices, X\=/ray imaging can be performed in~isolated rooms, so the~risk of~infection is significantly decreased \cite{Jacobi2020}. Nonetheless, especially on~X\=/ray images, it~is~particularly difficult to assess the~severity of~the~pathology, and, thus, only experts in~radiology should interpret chest images. In general, this process is faster than waiting for~RT\=/PCR test results. However, after individual patient collection, multiple test samples can be examined by a laboratory assistant simultaneously. Whereas a radiology technician is only able to take an image of~one patient at a time, and then such an image must still be analyzed by a radiologist.

Recent applications of~machine learning (ML) have gained popularity in~the medical domain \cite{Hu2018, Ayesha2021}. The~performance achieved by~neural networks is becoming similar to that reached by medical experts. Deep learning techniques for~medical images are present in~classification (skin lesions \cite{Barata2021}, lung cancer\cite{Yang2019}), detection (arrhythmia\cite{Hannun2019}, breast cancer\cite{Cheng2010}, pneumonia\cite{Rajpurkar2017}, ADHD\cite{Jaiswal2016}), segmentation (lung\cite{Souza2019}, brain\cite{Yang2019}) and~imaging reconstruction (magnetic resonance \cite{Jeelani2018}, Single Photon Emission Computerized Tomography (SPECT) \cite{Chrysostomou2020}). 

Considering the~need for~a~highly accurate and~fast diagnosis process, artificial intelligence (AI) can play a~significant role in~automating the~detection of~COVID\=/19 cases.

AI solutions are frequently based on~complex, so-called black-box models \cite{Rudin2018}. For this~reason, it~is~difficult to tell what factors lead to a~particular model prediction. Such a~lack of~interpretability may be~dangerous, as~it may lead to biased results and~incorrect decisions in~real diagnostic procedures. Recent development in~the area of~Explainable Artificial Intelligence (XAI) shows the~importance of~model explanations, which help to avoid erroneous predictions. Nevertheless, surprisingly, in~the area of~COVID\=/19 image analysis, there are still only few results concerning the~use of~XAI for~lung image analysis.

In~this paper, we will summarize recent publications about lung imaging analysis (section II-III), and~show how explainable AI techniques have been used in~these solutions (section IV). We will confront these approaches with the~domain knowledge of~radiologists and~we will show how many of~the~assumptions about data, models or~explanations made in~many of~the~analyzed studies are not appropriate. Finally, we will construct a~checklist to help model developers assess whether they avoided the~most common errors. We~believe that this criticism, together with the~proposed checklist, will contribute to building better models not only for~the~diagnosis of~COVID\=/19 disease, but~also for~other applications based on~lung images.

\section{Methods}

\subsection{Literature search}
\label{sec:literature}

\begin{figure}[!ht]

\centering

    \begin{tikzpicture}[
        start chain=going below,
        mynode/.style = {
                draw, rectangle, align=center, text width=22mm,
                font=\footnotesize, inner sep=1.5ex, outer sep=0pt,
                node distance=24mm and 4mm,
                on chain},
        nodereferences/.style = {
                draw, rectangle, align=center, text width=26mm,
                font=\footnotesize, inner sep=1.5ex, outer sep=0pt,
                node distance=24mm and 4mm,
                on chain},
        nodeexcluded/.style = {
                draw, rectangle, align=center, text width=30mm,
                font=\footnotesize, inner sep=1.5ex, outer sep=0pt,
                node distance=24mm and 4mm,
                on chain},
        finalnode/.style = {
                draw, rectangle, align=center, text width=18mm,
                font=\footnotesize, inner sep=1.5ex, outer sep=0pt,
                node distance=10mm and 4mm,
                on chain},
        widenode/.style = {
                draw, rectangle, align=center, text width=38mm,
                font=\footnotesize, inner sep=1.0ex, outer sep=0pt,
                node distance=24mm and 4mm,
                on chain},
        mylabel_1/.style = {
                draw, rectangle, align=center, rounded corners, 
                font=\footnotesize\bfseries, inner sep=1.5ex, outer sep=0pt,
                fill=cyan!30, minimum height=30mm,
                node distance=4mm and 4mm,
                on chain},
        mylabel_2/.style = {
                draw, rectangle, align=center, rounded corners, 
                font=\footnotesize\bfseries, inner sep=1.5ex, outer sep=0pt,
                fill=cyan!30, minimum height=28mm,
                node distance=4mm and 4mm,
                on chain},
        mylabel_3/.style = {
                draw, rectangle, align=center, rounded corners, 
                font=\footnotesize\bfseries, inner sep=1.5ex, outer sep=0pt,
                fill=cyan!30, minimum height=24mm,
                node distance=4mm and 4mm,
                on chain},
        every join/.style = arrow,
        arrow/.style = {very thick,-stealth}
    ]
    \coordinate (tc);
    
    \node (n1a) [widenode, left=of tc]  {Records identified through human-curated database searching:\\ 
                                        ArXiv, IEEE, Google Scholar, PubMed, Science Direct, Scopus, Web of~Science\\(n=27)};

    \node (n1b) [nodereferences,right=of tc]    {Additional records identified through references\\(n=4)};

    \node (n2)  [mynode, below=of tc]   {Full-text articles assessed for~eligibility\\
                                        (n =  31)};

    \node (n3)  [finalnode,join]        {Studies included in~qualitative synthesis\\ (n =  25)};
    
    \node (n2r) [nodeexcluded,right=of n2]    {Records excluded:\\ 
                                        - Irrelevant scope\\
                                        - Without XAI parts\\
                                        (n = 6)};
    
    \draw[arrow] ([xshift=+13mm] n1a.south) coordinate (a)
                                          -- (a |- n2.north);
    \draw[arrow] ([xshift=-8mm] n1b.south) coordinate (b)
                                          -- (b |- n2.north);
    \draw[arrow] (n2) -- (n2r);

    \begin{scope}[node distance=7mm]
        \node[mylabel_1,below left=0mm and 3mm of n1a.north west] {\rotatebox{90}{Identification}};
        \node[mylabel_2]  {\rotatebox{90}{Eligibility}};
        \node[mylabel_3]  {\rotatebox{90}{Included}};
    \end{scope}
    \end{tikzpicture}

\caption{PRISMA Flow Diagram shows the~flow of~information through the~different phases of~a~systematic review including inclusions and~exclusions.} 
\label{fig:Prisma}
\end{figure}
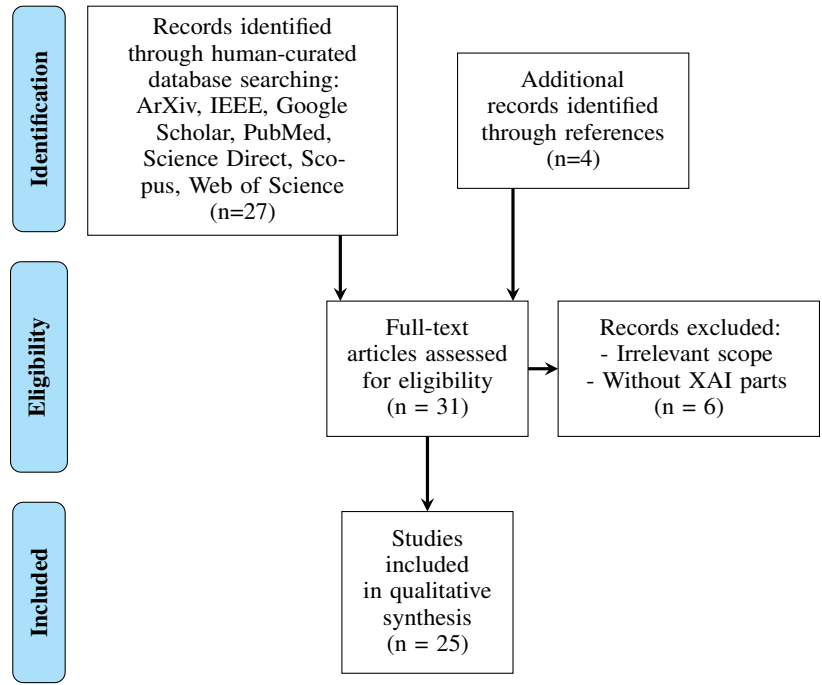

This research is based on~a~systematic literature review. The~data was collected between the~1st and~14th~of~August 2020. The~search was performed according to Preferred Reporting Items for~Systematic reviews and~the~Meta-Analyses (PRISMA) statement \cite{Liberati2009} presented in~Figure \ref{fig:Prisma} in~the following academic digital databases: ArXiv, IEEE, Google Scholar, PubMed, Science Direct, Scopus, Web of~Science. All studies written in~English, regardless of~the~publication status (preprint, peer-reviewed, or~published articles), were included in~this review. Studies were identified by~the~combination of~keywords: (“XAI COVID\=/19” OR~“explainable artificial intelligence COVID\=/19” OR~“explainable COVID\=/19” OR~“explanations COVID\=/19” OR~“interpretable COVID\=/19” OR~“interpretations COVID\=/19” OR~“transparent COVID\=/19”) and~(“X\=/ray” OR~“radiography” OR~“CT” OR~“computed tomography”). Then, each study was screened for~content relevance.

From~31 collected works, during eligibility checking, 6~studies were dropped due to irrelevant scope or~lack of~XAI parts. There were 25 studies included for~qualitative synthesis. The~number of~studies considered in~the review is vast enough to create a~representative set/collection for~further investigation. Some of~the~studies were published as preprints, not as camera-ready articles. Due to the~rapidly growing field of~tools related to supporting medical practitioners in the~fight against pandemics, we included them. They will help to show a~variety of~considered XAI approaches.

Then, on February 26, 2021, it was verified which articles previously available as preprints were published in peer reviewed journals. The content of~these articles was again reviewed for changes that had been made. Articles published in journals: \cite{b8, b13, b16, b18, b19, b20, b21, b26, b27, b28, b29, b30}, preprints: \cite{b5, b6, b7, b9, b10, b11, b12, b15, b17, b23, b24, b25, b31}.

\subsection{Types of~COVID-19 prediction tasks}

\begin{figure}[!ht]
    \centering
    \begin{tikzpicture}[sibling distance=.3cm,scale=1.0]
    \tikzset{
        edge from parent/.style= 
            {thick, draw, edge from parent fork right},
        level 1/.style={draw, rotate=90,minimum width=2.3cm, text width=2.8cm, align=center, level distance=3.2cm},
        level 2+/.style={draw,minimum width=4.1cm, text width=4.0cm, align=center, level distance=5.0cm},
        every tree node/.style={draw},
        grow'=right}
    \Tree 
    [. \rotatebox{90}{ COVID-19 problems related to medical imaging in~the~reviewed studies} 
        [.{Segmentation}
            [.{Lung \\ \cite{b19, b20,  b26, b12, b25}} ]
            [.{Lesion \\ \cite{b20, b30, b17}} ]
        ]
        [.{Classification \\(COVID-19 vs.~other classes)}
            [.{Binary \\ \cite{b8, b13, b18, b21, b27, b30, b5, b7, b12, b15, b23, b24}} ]
            [.{Three-class \\ \cite{b13, b16, b21, b27, b28, b29, b17, b23, b31}} ]
            [.{Four-class \\ \cite{b19, b27, b11}} ]
            [.{Five-class \\ \cite{b6}} ]
            [.{Multi-label \\ \cite{b9}} ]
        ] 
        [.{Classification \\ (severity assessment)} 
            [.{Parts of~lungs rating \\ \cite{b26, b25}} ]
            [.{Multi-label lesions \\ \cite{b10}} ]
        ]
    ]
   \end{tikzpicture}
    \caption{Taxonomy of~AI applications in~25 reviewed studies}
    \label{fig:taxonomy}
\end{figure}
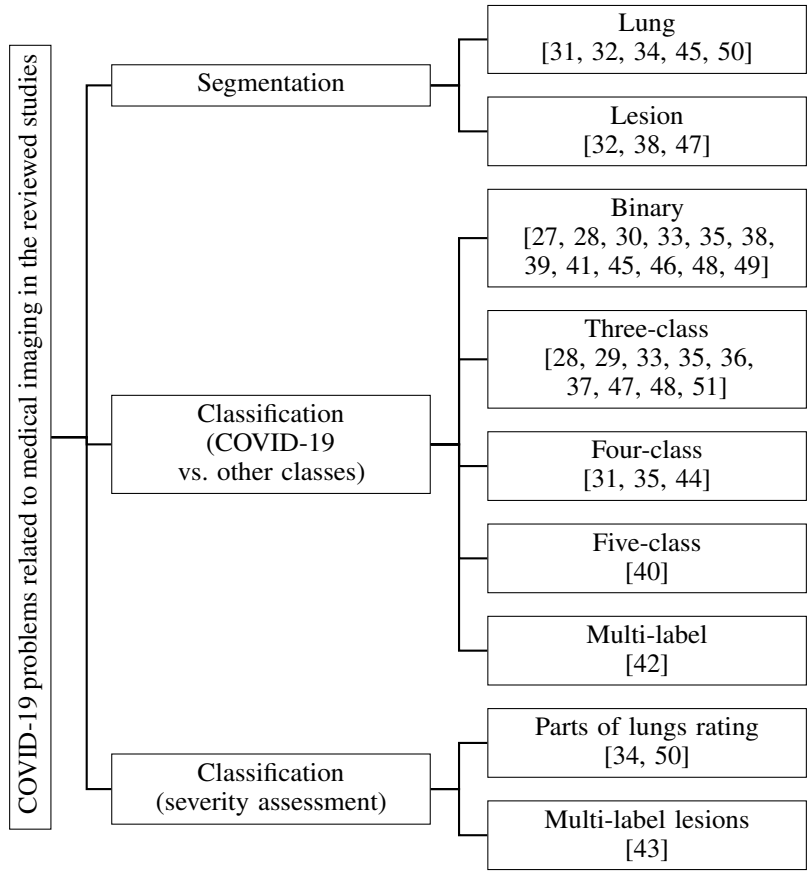

The 25 studies identified according to our methodology concern various types of~ML tasks.
In~Figure \ref{fig:taxonomy}, we introduce a~taxonomy for~these studies. In~the~following sections, when discussing specific solutions, it~is~important to remember what kind of~problems they are designed for.

The first breakdown concerns the~types of~tasks, such as: classification (COVID\=/19 vs. other classes), severity assessment or~segmentation.
Among classification tasks, the~first group is related to the~detection of~COVID\=/19 cases. The~goal of~the~second is to assess how severe the~changes caused by~COVID\=/19 are.

In~the~reviewed works, segmentation can be regarded as an image preparation technique for~the~further classification process. Segmentation was not used in all studies. Lung segmentation is present in~works \cite{b19, b20, b26, b12, b25} and~lesion segmentation in~\cite{b20, b30, b17}. The~lungs are segmented to remove the~unnecessary background because, based on~medical experience, the~lesions caused by~COVID\=/19 are not located outside the~lungs. Lesion segmentation, also called infection mask, helps to train the~model to recognize infected regions and~can be beneficial for~further model assessment. Only radiologists can prepare trustworthy and~high-quality lung lesion masks, not an automatic tool similar to those in~the~reviewed works \cite{b20, b30}. However, manual mask preparation takes a~lot of~time and~money and~requires a~high level of~consistency among annotators, but, surely, it is the~most valuable method for~qualitative and~quantitative XAI evaluation.

There are not any strict guidelines on~how many classes the~classification should be conducted on. In~classification problems, the~number of~classification classes varies between studies. This distinction is~particularly important when comparing model performance. It is worth noting that some studies have verified the~performance of~the model considering different number of~classes \cite{b13, b21, b27, b23}. Binary classification task detecting COVID\=/19 and~non\=/COVID\=/19 is the~most popular \cite{b8, b13, b18, b21, b27, b30, b5, b7, b12, b15, b23, b24}. Another frequently used method is three-class classification \cite{b13, b16, b21, b27, b28,  b29, b17, b23, b31}: no infection/no pneumonia, pneumonia (bacterial or~non\=/COVID\=/19 viral infection), and~COVID\=/19 (COVID\=/19 viral infection). In~the~four-class classification problem, there are two different approaches to~splitting images. The first one divides images into: normal, bacterial pneumonia, non\=/COVID\=/19 viral pneumonia, and~COVID\=/19 \cite{b27, b11}, while the~second one into: normal, bacterial pneumonia, tuberculosis (TB), and~viral pneumonia. In~the~second approach, study \cite{b19} refers to \cite{Yoon2020} and~claims that differentiation between viral pneumonia and~COVID\=/19 is challenging, because similar radiological features exist between them. In~classification tasks with the~biggest number of~classes (5), the~authors distinguish five different classes: normal, pneumonia, virus, bacteria, and~COVID\=/19 \cite{b6}. In~\cite{b9} the~authors applied multi-label classification. For~example, for~an image with the~lungs of~a~patient suffering from~COVID\=/19, the~correct classification should predict all three labels: pneumonia, viral pneumonia, and~COVID\=/19.

In~the~segmentation for~severity assessment, there are two different approaches. One rates how severe the~lesions are in~parts of~the~lungs. Another one assigns labels to lesions' names that point out which changes in~the lungs are present in~the image.

In~two studies \cite{b26, b25} images or~parts of~images were divided into classes that correspond to the~severity of~COVID\=/19 effects on~the~lungs. The~authors \cite{b26} propose dividing each lung horizontally into 4 parts, and~giving them a~grade of~1 if~it contained any lesions, such as consolidation or~ground-glass opacities, or~0 otherwise. Then, the~grades from~all parts were summed up. Based on~grades, the~following scale was prepared: Normal\=/PCR+ 0 (a patient classified by expert radiologist as Normal that has positive RT\=/PCR test), Mild 1\=/2, Moderate 3\=/5, and~Severe 6\=/8. A similar solution was introduced by~\cite{b25}. In~this study, each lung was divided horizontally into 3 parts and~a~4 grade scale was used for~each part. The~division of~each lung into 3 parts (upper, middle, and~lower lung field) resembles natural lung structure and~common radiological practice.

An example of~multi-label classification is described in~\cite{b10} where 18 outputs of~neural network (such as atelectasis, consolidation, infiltration, pneumothorax, edema, emphysema, fibrosis, effusion, pneumonia, pleural thickening, cardiomegaly, nodule, mass, hernia, lung lesion, fracture, lung opacity, and~enlarged cardiomediastinum) give information about lesions in~lungs. Some of~them are characteristic for~specific illnesses. This is a~very promising approach because with XAI visualizations it should be clearly visible for~radiologists if~a~model learned to recognize proper lesions.

\section{Deep learning on lung image data}

\subsection{Data resources} \label{Data_resources}

The COVID\=/19 virus is a~relatively new disease, and, in~several articles, the~lack of~high-quality medical imaging databases is indicated \cite{b26, b17}. 
To verify this, we looked at the~quality of~the data used to construct the~models in selected peer-reviewed research descriptions. The results of~this analysis are presented in~Table~\ref{tab:links}.

For medical imaging, the~standard format for representing measured and / or~reconstructed data is~DICOM (Digital Imaging and Communications in Medicine). Significant features of~this format are~the~ability to faithfully record 16-bit dynamics of~data in grayscale (CT, radiography), control of~acquisition parameters, and the~ability to adapt to presentation conditions at diagnostic stations. The use of~full data dynamics, i.e., all information about the~imaged objects, taking into account the~characteristic properties of~the~entire measurement and reconstruction process of~the data (equipment parameters, filters, parameters of~the acquisition process, specificity of~pre-processing and forms of~image representation) enables the~construction of~models based on complete measurement information about the~examined object. Unfortunately, the~vast majority of~COVID\=/19 resources do not retain the~source image information on the~diagnosed objects. The~data is converted from DICOM to typical multimedia image formats (mainly JPEG, PNG, TIFF standards) with the~omission of~information about the imaging process itself and often with a loss of~quality and informative value of~the compressed data. Data dynamics is often reduced to the 8 most significant bits, quantized to simplified 8-bits representation or~all image information is lossy compressed (JPEG) using standard quantization tables.

{
\footnotesize
\setlength{\tabcolsep}{3pt}

\begin{longtable}{|p{0.015\textwidth}|p{0.1\textwidth}|p{0.15\textwidth}|p{0.07\textwidth}|p{0.10\textwidth}|p{0.10\textwidth}|p{0.415\textwidth}|}

\caption{This table presents the~data sources used in~studies reviewed in~this survey. For~each data source, we list articles that use it. The JPEG quality factor (QF) for most images has been set to 75, other cases are indicated. In~the~case of~COVID\=/Net, please note that it is not a~data source, but~a~study collating 5 datasets. Some other studies refer to it instead of~referring to~the~original source.}
\label{tab:links}

\\ \hline
\textbf{No.} & \textbf{Institution} & \textbf{Link to dataset}  & \textbf{Used in~article}      & \textbf{Dynamic range of images} & \textbf{Data processing} & \textbf{Prepared for scientific experiments}\\ \hline

1) & University of~Waterloo  & \url{github.com/lindawangg/COVID-Net}                       & \cite{b29, b17, b23, b25}     & \multicolumn{3}{|l|}{}  \\ \hline
2) & \hangpara{5mm}{0}University of~Waterloo & \url{github.com/agchung/Figure1-COVID-chestxray-dataset}       & \cite{b16, b23}               & 8 bits, 48 cases & JPG, PNG & X-ray database for research purposes only, continuously growing; Metadata: offset, sex, age. finding, survival temperature, pO2, saturation, view, modality, artifacts/distortion, notes; Categories: covid, pneumonia, no finding \\ \hline
3) & \hangpara{5mm}{0}University of~Waterloo & \url{github.com/agchung/Actualmed-COVID-chestxray-dataset}    &                               & 8 bits, 237 cases & PNG, BMP & X-ray database for  research purposes only, continuously growing; Metadata: finding, view, modality, notes; Categories: covid, no finding \\ \hline
4) & \hangpara{5mm}{0}Qatar \& Bangladesh Universities  & \url{kaggle.com/tawsifurrahman/covid19-radiography-database} & \cite{b16}                    &8 bits, 21165 cases&PNG, resized&X-ray database; No metadata; Categories: COVID-19 positive cases (3616),  normal (10,192), lung opacity (Non-COVID lung infection - 6,012), viral pneumonia (1,345) \\ \hline
5) & \hangpara{5mm}{0}University of~Montreal  & \url{github.com/ieee8023/covid-chestxray-dataset}  & \cite{b8, b16, b18, b19, b21, b27, b28, b6, b9, b10, b11, b15, b23, b25, b31} & 8 bits, 951 cases &JPG, PNG, resized&X-ray database; Metadata: covid severity scores, sex,age, finding, RT\_PCR\_positive, survival, intubated, intubation\_present, went\_icu, in\_icu, needed\_supplemental\_O2, extubated, temperature, pO2\_saturation, leukocyte\_count, neutrophil\_count, lymphocyte\_count, clinical\_notes, other\_notes; Categories: covid, viral, bacterial, fungal, lipoid, aspiration, unknown \\ \hline
6) & \hangpara{5mm}{0}National Institutes  of~Health & \url{kaggle.com/c/rsna-pneumonia-detection-challenge}   & \cite{b16, b10, b23, b25}     &8 bits, 30227 (training)+3000 (test) cases&DICOM, resized&X-ray database of Pneumonia Detection Challenge; No metadata; Categories: normal. lung opacity, no lung opacity/not normal  \\ \hline
7) & National Institutes of~Health  & \url{nihcc.app.box.com/v/ChestXray-NIHCC}                   & \cite{b21, b10, b23}          &8 bits, 112120 cases&PNG, resized&X-ray database of Common Thorax Disease; Metadata: finding ROI; Categories: no findings and 14 disease categories (Atelectasis, Cardiomegaly, Effusion, Infiltration, Mass, Nodule, Pneumonia, Pneumothorax, Consolidation, Edema, Emphysema, Fibrosis, Pleural\_Thickening, Hernia)  \\ \hline
8) & National Institutes of~Health & \url{kaggle.com/nih-chest-xrays/sample}                     & \cite{b8}     &8 bits, Random sample of 5606 from 112,120 images of 30,805 unique patients&PNG, resized&X-ray database; Metadata: finding labels, follow-up, age, gender, view; Categories: Atelectasis, Cardiomegaly, Effusion, Infiltration, Mass, Nodule, Pneumonia, Pleural\_Thickening, Hernia, Pneumothorax, Consolidation, Edema, Emphysema, Fibrosis \\ \hline
9) & University of~Montreal   & \url{kaggle.com/praveengovi/coronahack-chest-xraydataset} & \cite{b19, b6}                     &8 bits, 5910 cases (normal-1576, covid 58, SARS-4, virus-1493, bacteria 2777, ARDS-2)&JPG,PNG-resized&Collection Chest X Ray (anterior-posterior) of Healthy vs Pneumonia (Corona) affected patients infected patients along with few other categories such as SARS (Severe Acute Respiratory Syndrome ), Streptococcus \& ARDS (Acute Respiratory Distress Syndrome); No metadata  \\ \hline
10) & University of~California San Diego & \url{kaggle.com/paultimothymooney/chest-xray-pneumonia}       & \cite{b19, b27, b9, b11, b31} &8 bits, 5863 cases&JPG&Chest X-ray images (anterior-posterior) were selected from retrospective cohorts of pediatric patients of one to five years old from Guangzhou Women and Children’s Medical Center, Guangzhou. All chest X-ray imaging was performed as part of patients’ routine clinical care.; Categories: normal and pneumonia; No metadata \\ \hline
11) & University of~California San Diego & \url{github.com/UCSD-AI4H/COVID-CT}                       & \cite{b18, b5, b7, b15}       &8 bits, 349 cases&Images collected (scanned) from covid-related and medical papers in PNG (covid) or JPG (normal)&This dataset has 349 CT images containing clinical findings of~COVID-19 from 216 patients; Categories: covid and noncovid cases; Metadata: age, gender, location, medical history (unfortunately modest), time after the onset of illness, severity, other diseases  \\ \hline
12) & University of~California San Diego  & \url{data.mendeley.com/datasets/rscbjbr9sj/2}  & \cite{b28}                   &8 bits, 5233 cases&JPG (QF=95 for normal and QF=75 for pneumonia)&Collection Chest X Ray; Categories: normal (1349 cases) vs pneumonia (3884 cases) including subcategories of bacteria and virus; No metadata \\ \hline
13) & Elazig in Turkey & \url{github.com/muhammedtalo/COVID-19}                   & \cite{b8, b16}                &8 bits, 1125 cases&JPG (QF=90, subsampling2x2), PNG (resized)&X-Ray Images collection; No metadata; Categories: covid (125 cases), no findings (500 cases), pneumonia (500 cases) \\ \hline
14) & National Library of~Medicine & \url{openi.nlm.nih.gov/gridquery?it=xg&coll=cxr&m=1&n=100}                                      & \cite{b19, b10}              &8 bits or full bits, 7470 cases&PNG (resized), Full DICOM&Chest X-rays collection with 3,955 radiology reports; Categories: 14 pulmonary categories; Metadata: time after the onset of illness, severity, other diseases, captions of symptoms as unstructured symptom description\\ \hline
15) & Stanford University School of~Medicine & \url{stanfordmlgroup.github.io/competitions/chexpert}  & \cite{b10, b25}               &8 bits, 224,316 chest radiographs of 65,240 patients&JPG&Large dataset of chest X-rays which features uncertainty labels and radiologist-labeled reference standard evaluation sets; Categories: each report was labeled for the presence of 14 observations (no finding, enlarged cardiom., cardiomegaly, lesion, opacity, edema, consolidation, pneumonia, atelectasis, pneumothorax, pleural effusion, pleural other, fracture, support devices) as positive, negative, or uncertain; Metadata: related to above categories (blank for unmentioned, 0 for negative, -1 for uncertain, and 1 for positive) \\ \hline
16) & Hospital San Juan de Alicante - University of~Alicante & \url{bimcv.cipf.es/bimcv-projects/padchest}              & \cite{b10}                   &8 bits, more than 160,000 images from 67,000 patients&PNG&PadChest: A large chest x-ray image dataset with multi-label annotated reports; the reports were labeled with 174 different radiographic findings, 19 differential diagnoses, and 104 anatomic locations; a 27\% of the reports were manually annotated by trained physicians; Metadata: age, sex  \\ \hline
17) & Hospital Universitario San Cecilio & \url{github.com/ari-dasci/OD-covidgr}                        & \cite{b26}                    &8 bits, 852 images&JPEG (QF=90)&X-ray images: 426 positive covid cases and 426 negative cases; only the posterior-anterior view is considered; Categories: covid severity - normal-PCR+ (76), mild (100), moderate (171), severe (79); General metadata: positive images correspond to patients who have been tested positive with RT-PCR within a time span of at most 24h between the X-ray image and the test; every image has been taken using the same type of equipment and with the same format  \\ \hline
18) & Beth Israel Deaconess Medical Center in~Boston & \url{physionet.org/content/mimic-cxr/2.0.0} & \cite{b10}                    &full bits, 227,835 imaging studies for 65,379 patients&full DICOM&Chest radiographs with metadata: electronic health record data, dicom metadata, free-text radiology reports Categories: 14 pulmonary observations with an additional “uncertain” category \\ \hline
19) & Società Italiana di Radiologia Medica e Interventistica & \url{sirm.org/category/senza-categoria/covid-19}              & \cite{b24}                    &8 bits&mostly JPG (QF=95, subsampling2x2)&Chest radiographs with free-text radiology and clinical reports, covid confirmation; Metadata includes selected information from electronic health record (e.g. symptoms, lab exams, ARDS, ventilatory assistance, previous exams); Categories: covid confirmation or no with 14 pulmonary observations  \\ \hline
20) & National Cancer Institute  & \url{wiki.cancerimagingarchive.net/display/Public/LIDC-IDRI} & \cite{b12}                    &full bits, 1308 cases&full DICOM&The Lung Image Database consists of diagnostic and lung cancer screening thoracic CT scans with marked-up annotated lesions (XML); it includes three categories ("nodule > or =3 mm," "nodule <3 mm," and "non-nodule > or =3 mm");  \\ \hline
21) & University of Brescia     &\url{brixia.github.io/#dataset}                     & \cite{b25}   & full bits, 4,707 cases & full DICOM & COVID-19 subjects, acquired with both CR and DX modalities, in~AP~or~PA projection with highly expressive multi-zone COVID-19 severity score, fully annotated; Metadata: the multi-region 6-valued Brixia-score defined for six zones, sex, age \\ \hline

22) & open-edit radiology resource & \url{radiopaedia.org}                                         & \cite{b24}                 &8 bits, a significant number of cases, constantly updated&JPG with different QF, resized&Database of general radiological purposes; in selected cases free-text radiology and clinical reports, selected; generally, quantitatively and qualitatively differentiated case reports \\ \hline
23) & generated using data augmentation & \url{kaggle.com/nabeelsajid917/covid-19-x-ray-10000-images}   & \cite{b16}                    &8 bits, 104 cases&JPEG with different QF, resized&Corona Virus X-ray Dataset; Categories: covid and normal; No~metadata  \\ \hline

24) & \multicolumn{2}{|l|}{offline database or~from~hospital}            & \cite{b13, b20, b30, b12}     & \multicolumn{3}{|l|}{}  \\ \hline

\end{longtable}
}

\paragraph{\textbf{Scarcity of~publicly available COVID\=/19 data sources with images in~raw~DICOM format}} It~was observed that only one out of~five repositories with~the~DICOM extension presented in~Table \ref{tab:links} contains COVID\=/19 cases. Most databases regarding COVID\=/19 images are in 8-bit~JPG or~PNG formats. There are~concerns that the quality of~shared images is degraded, which may render the trained models less accurate. The quality degradation includes: the Hounsfield unit (HU) values are inaccurately converted into grayscale image data, and the number of~bits per pixel and the resolution of~images are reduced.

An extreme case is the use of~digital scans of~printed lung images with no standard regarding image size, e.g., images extracted from the manuscripts \cite{b24}. Comparative statistical analysis based on the value of~systematic measurement errors for the COVID-19 data, including the raw data and the metadata extracted from official reports, showed noticeable and increasing measurement errors \cite{Barata2021}. This matter showed~the importance of~the accuracy, timeliness and completeness of~COVID-19 datasets for better modeling and~interpretation.

\paragraph{\textbf{Too few images with low and~moderate severity cases}} Most studies are based on~data sources publicly available on~the~Internet on~a~popular sharing platform, such as GitHub and~Kaggle. The~most commonly used data source with COVID\=/19 cases was created at the~beginning of~the~epidemic. The~first publicly available repository was published on~January 15, 2020. In~\cite{b26} it is stressed that available data sources contain too few images with low and~moderate severity cases. Most of~the~data sources have only class labels without fine-grained pixel-level annotations, for example 3), 4), 8), 9), 10).

\begin{table*}[ht]
\centering
\caption{Class balance in~the~reviewed studies. The~class balance is crucial developing an accurate model. In~the~following rows there are presented: the~number of~COVID\=/19 images in~the~study, the~total number of~images, the~number of~classes into which images were divided, and~aspect ratio. The~aspect ratio is calculated by~dividing the~number of~COVID\=/19 images by~the~total number of~images and~then multiplying it by~the~number of~classes. The~biggest collected COVID\=/19 dataset and~the~largest total number of~images are marked in~bold. The~smaller the~aspect ratio, the~less COVID\=/19 cases participate in~the whole study's dataset, and~vice versa. For~this reason, the~best-balanced dataset (nearest 1) is marked in bold. Studies which do not include full information about the~number of~cases were excluded.}

\label{tab:balance}

\newcolumntype{L}[1]{>{\raggedright\let\newline\\\arraybackslash\hspace{0pt}}m{#1}}
\def\arraystretch{1.4}
\setlength{\tabcolsep}{0.6pt}

    \scalebox{0.9}{
    \begin{tabular}{|L{2.1cm}|c|c|c|c|c|c|c|c|c|c|c|c|c|c|c|c|c|c|c|c|c|c|c}
    
    \hline
    \textbf{Study}                     & \cite{b8} & \cite{b13} & \cite{b16} & \cite{b18} & \cite{b19} & \cite{b20} & \cite{b21} & \cite{b26} & \cite{b27} & \cite{b28} & \cite{b29} & \cite{b30} & \cite{b5} & \cite{b6} & \cite{b9} & \cite{b11} & \cite{b12} & \cite{b17} & \cite{b24} & \cite{b31} \\ \hline
    \textbf{Number of~COVID\=/19 images} &250&230& \textbf{855} &345&502& \textbf{3,389} &127&377& 112/137 &76&358&400&400&58&234&68& 829$^{*2}$ &99&120& 269 \\ \hline
    \textbf{Total number of~images}    &6523&460& \textbf{15,959} &720&1004&2186&1125&754&366&5949& \textbf{13,975} &750&800&2800&1234&5941& 1,865$^{*2}$ & \textbf{18,529} &239& 5 801 \\ \hline
    \textbf{Number of~classes}         & 2/3 &3&3&2&4&2&3& $\frac{5}{4}^{*3}$ & 3/2 &3&3&2&2&4& 12$^{*1}$  &4&2&3&2& 3 \\ \hline
    \textbf{Aspect ratio}              & 0.08/0.11 &1.5&0.16& \textbf{0.96} &2&1.36&0.34&0.63& 0.92/0.75 &0.04&0.08&1.07& \textbf{1.00} &0.08&2.28&0.05&0.89&0.02& \textbf{1.00} & 0.14 \\ \hline
    
    \end{tabular}
    }
    \\
    
\raggedright

$^{*1}$ multilabel classification, $^{*2}$ training slides (106 COVID-19 images), $^{*3}$ four COVID\=/19 classes (NormalPCR+: 76 cases, Mild cases: 80, Moderate: 145 cases, Severe: 76 cases) and~one Negative (377~cases)

\arrayrulecolor{black}
\end{table*}

\paragraph{\textbf{Relatively low number of~COVID\=/19 images}} Image format is one problem, while the~amount of~available data is another problem. The~median number of~COVID\=/19 images in~the~considered data resources is $250$. With so little data, it is difficult to train a~deep neural network (DNN). Table \ref{tab:balance} shows the~number of~cases in~particular classes. The~last row with aspect ratio shows the~proportion of~the~COVID cases to~non\=/COVID cases.

The~use of~imbalanced datasets requires more attention during the~model training. Either proper data resampling \cite{b20} (oversampling \cite{b28, b23}, undersampling \cite{b27}) should be applied, or~an~appropriate loss function should be chosen \cite{b16, b11, b15, b17}, unless acquiring a~greater amount of~less common data is possible. It is also possible to use micro-base metrics \cite{b9}. However, most ML algorithms do not work very well with imbalanced datasets.

\paragraph{\textbf{The data sources lack descriptions}} Data resources: 4), 6), 9), 10), 12), 13), and 23) did not include metadata. At~a~minimum, the~description of~the~dataset should include the~following factors. First of~all, the~total number of~images and~the number of~images in~each class should be given. Additionally, the balance in~terms of~age and~sex is another important factor because of~the~differences in~anatomy. Information about smokers or~previous lung diseases is also relevant. For~analyzing model responses, the~information about concurrent diseases, the~severity of~COVID\=/19, and~the number of~days between the~exposure and~the~acquisition of~the~image of~the~chest are also useful.

\paragraph{\textbf{Mix of~CT and~X\=/ray images}} The~problem that we found in~these datasets is the~data purity.
If we look closer at~the~images presented in~study \cite{b15}, it appears that CT and~X\=/ray images are mixed in~the~X\=/ray dataset. These two techniques are so different that networks for~CT and~X\=/ray images should be trained separately.

\paragraph{\textbf{Inappropriate CT windows}} For~COVID-related lung analysis, it is essential to have Hounsfield Units equivalent for~“lung” window (width: 1,500, level: -600). Otherwise, the~lung structures are obscured or~not visible at all, such as some examples in~studies \cite{b13, b5, b7}. This is a~basic, but~key, issue because we do not want to assess soft tissues or~bones. Photos taken in~other windows do not have any real diagnostic value.

\paragraph{\textbf{“Children are not small adults”}} When we go back to the~databases, it appears that, in~some cases, e.g.~7), 8), 9) the X\=/rays of~children and~adults are mixed. The next problem is related to the~mixture of~images of~patients of~different ages. There are crucial differences between the X\=/rays of~children's and~adults' chests: technical (hands are often located above the~head), anatomical (different shape of~the~heart and~mediastinum, as well as bone structures), and~pathological (different pathologies). This will also include a~different course of~infectious diseases, with the~most vivid example of~round pneumonia mimicking tumors \cite{Lee2020}. It is important to mark the age of~patients in data resources, and to separate children from adults when preparing data for training.

\paragraph{\textbf{CT and X-rays images are not in color.}} Despite that fact, some databases, e.g. 5) and 11), include images in RGB color space. It introduces redundant information, because values in all channel are the same (R=G=B). This situation leads to increasing the number of~input neurons in the neural network by three times. Due to that fact, the number of~parameters will also rise, and the training may require more data and time, however, it lacks extra information.

\paragraph{\textbf{Incorrect categorization of~pathologies}} We have noticed that some images are incorrectly categorized - into normal or~pathologic, e.g. in database 10), 13), and~also within the~class of~pathology, e.g. in database 14). An~additional problem is that, from~a~medical point of~view, some images should be multi-categorized. This means that there is more than one pathology in~one image. For~instance, pneumonia (main class) can manifest itself as lung consolidations, which, however, can also appear with pleural effusion or~atelectasis (two additional classes). On the~other hand, atelectasis itself, with a~mediastinal shift, can be a~sign of~a~different pathology, such as a~lung tumor. Thus, databases should be verified by~experienced radiologists for~proper categorization and~maybe a~rejection of~multi-class images. This, however, would be time-consuming and~- what is more important - very difficult or~impossible with low-quality images or~images without appropriate descriptions.

\begin{figure}[!ht]
    \centering
    \begin{tabular}{p{0.5\textwidth} p{0.5\textwidth}}
      \includegraphics[width=.5\textwidth]{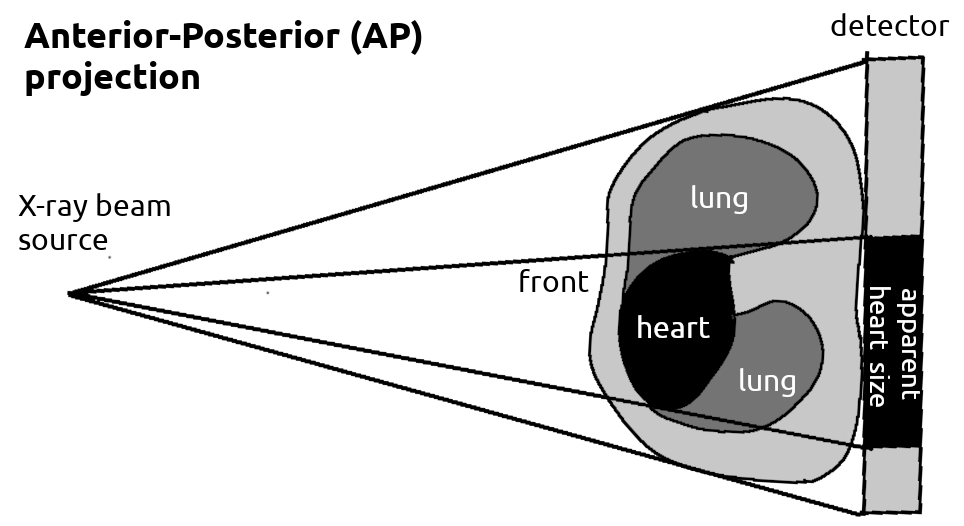} &
      \includegraphics[width=.5\textwidth]{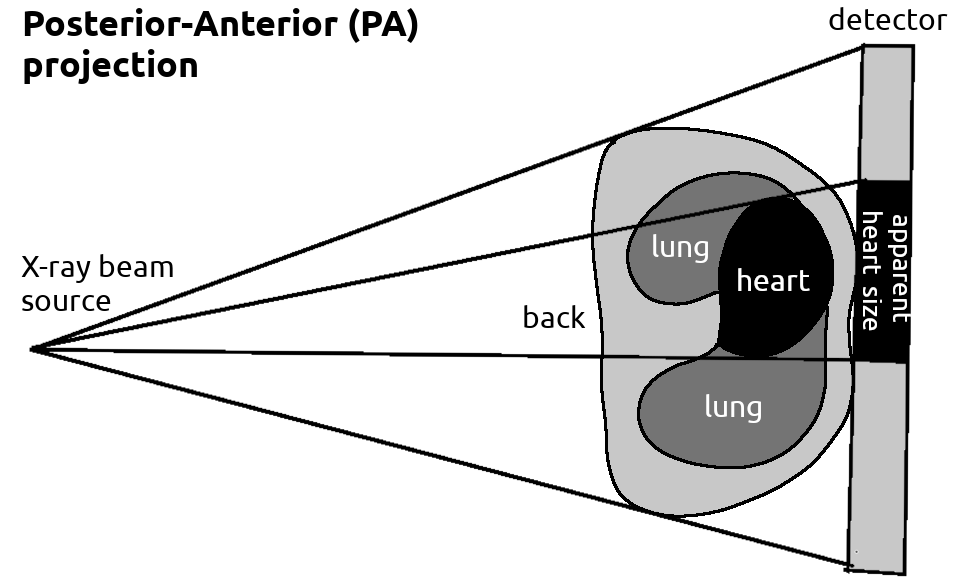}
    \end{tabular}
    \caption{Differences between AP and~PA chest projections}
    \label{fig:AP_PA}
\end{figure}

\paragraph{\textbf{Lack of~information about chest projection for~X\=/ray imaging}} This problem is present, for example in 2), 4), 9).  There are two main chest projections, see Figure \ref{fig:AP_PA}, Posterior-Anterior (PA) and~Anterior-Posterior (AP). The~first one is acquired while the~patient is standing. The~X\=/ray beam comes through the~patient's chest from~its back (posterior) to front (anterior) - i.e., PA projection. The~second one is the~opposite - the~beam enters through the~front (anterior) of~the~chest and~exits out of~the~back (posterior) - i.e., AP projection. This type of~examination is mostly conducted in~more severe cases, with~lying patients, with comorbidities, often in~Intensive Care Units. As the~X\=/ray beam is cone-shaped, both projections have one very important difference, which is the~size of~the~heart. In~PA projection, the~heart is close to~the~detector, resulting in~a~similar heart size on~the~X\=/ray as in~reality. In~AP projection, the~heart is away from~the~detector, resulting in~a~larger heart size on~the~X\=/ray, which can be confused with cardiomegaly. In~databases, AP and~PA images are often mixed, which can cause bias because AP projections are performed on~severely ill patients \cite{b26}. From~a~medical point of~view, it is impossible to perform chest X\=/rays in~only one projection as this depends on~patients condition. However, projection should be specified for~every X\=/ray in~dataset, and~possible bias in~model classification should be evaluated.

To sum up, this section shows that data sources have several weaknesses. First of~all, images available for~COVID\=/19 in~public databases are in~not the most dedicated image format because DICOM images are still rarely available for~this disease. Secondly, in~the~data sources, there is missing data (i.e., chest projection) or~poor quality data (i.e., poor image quality, not grayscale images, inappropriate CT window, mixed CT and~X\=/ray images, or~incorrect pathology categorization). Lastly, during data preparation, it~should be taken into account that DNNs work better when the~class balance is maintained.

\subsection{Image preprocessing methods}

The aim of~preprocessing is to make the~images from~different data resources look homogeneous and~coherent. This process reduces the~possibility of~bias via eliminating some artifacts from~images, such as captions, annotations, which may deceive the model. The~model should learn how to differentiate labels by~focusing on~image features, not by~recognizing from~which database the~image comes from. During preprocessing, irrelevant image features that are easier to learn are removed. This is because in~some databases there are no cases of~people suffering from~COVID\=/19, while in~others there are, for~example, only serious cases. These differences, which are insignificant from~a~human perspective, must be eliminated. For~machines, even the~information that images from~one data resource are relatively darker might be relevant.

However, due to a~large amount of~data, automation of~preprocessing is necessary. Preprocessing cannot introduce any changes in~an image which will add or~remove some relevant information. Its purpose is~to~make it impossible to identify the~machine or~characteristic machine's calibration parameters, e.g.,~the~dose of~exposure.

Table \ref{tab:preprocessing} lists preprocessing techniques used in~the~reviewed studies. The~most common was resizing an~image to the~same size. It is the~most basic operation needed to train DNN when images have different sizes. Other techniques applied frequently to images were: normalizing pixel intensity, changing color space, eliminating noise, equalizing histogram, and~performing image enhancement. Unfortunately, in~9 out of~25 studies there was no information about preprocessing steps provided. \cite{Wynants2020} also stressed that many studies do not contain sufficient information about preprocessing, such as cropping of~images.

\begin{table}[!ht]
\caption{Image preprocessing techniques in~the~reviewed studies}
\label{tab:preprocessing}
\centering
\begin{tabular}{|>{\raggedright}l|l|}
\hline
\textbf{Preprocessing technique}       & \textbf{Reference}                                         \\ \hline
Resize to the~same size                 & \cite{b16, b18, b19, b20, b27, b28, b6, b7, b9, b10, b11, b15, b24, b31} \\ \hline
Normalize pixel intensity               & \cite{b16, b27, b9, b10, b11, b24, b25}                     \\ \hline
Eliminate noise                         & \cite{b16, b18, b26, b25}                                   \\ \hline
\hspace{3mm} Use Perona-Malik filter    & \cite{b16}                                                  \\ \hline
\hspace{3mm} Limit image intensity      & \cite{b9, b25}                                              \\ \hline
Equalize histogram                      & \cite{b16, b19, b6, b25}                                    \\ \hline
Perform image enhancement               & \cite{b16, b18, b19, b20}                                   \\ \hline
Cast data type                          & \cite{b16, b19, b25}                                        \\ \hline
Change color space                      & \cite{b28, b15, b24}                                        \\ \hline
Crop image                              & \cite{b26, b6, b9}                                          \\ \hline
Zoom image / augmentation               & \cite{b7, b24}                                              \\ \hline
Add pixels                              & \cite{b26}                                                  \\ \hline
Feature encoding                        & \cite{b31}                                                  \\ \hline
Rotate image                            & \cite{b7}                                                   \\ \hline
Use 2D wavelet transform                & \cite{b18}                                                  \\ \hline
Feature extraction                      & \cite{b17,b31}                                              \\ \hline
Lack of~preprocessing or~description & \cite{b8, b13, b21, b29, b30, b5, b12, b17, b23}     \\ \hline
\end{tabular}
\end{table}

Cropping, changing color space, proportionally resizing, or~zooming can be helpful to adjust images for~training on~specific network architecture, or~the~easiest way to remove some descriptions from~the~edges of~images. If~not required, resizing ought to be omitted. Normalizing pixel intensity or~equalizing histograms are required to eliminate strong correlations with specific machine settings. Spot changes, such as noise removal, are not desirable. These techniques can be used  only very carefully in~order not to remove important features, such as lesions or~parts of~them.

To sum up, preprocessing is an important step preceding model training. It should reduce the~possibility of~bias and~guarantee more homogeneous images without the~elimination of~any medically significant features.

\subsection{Data augmentation}

Data augmentation for~ML is a~technique that artificially multiplies the~number of~images through cropping and~transforming existing images or~creating new synthetic images thanks to generative adversarial networks. This procedure may help to reduce model overfitting and~the problem of~class imbalance. It helps in~achieving a~larger training dataset and~more robust models.

In~Table \ref{tab:augmentation}, we summarized data augmentation techniques from~the~reviewed studies. The~most popular augmentation techniques in~the~reviewed studies are affine transformations, such as rotation, scaling or~zooming, flip, and~shifting or~translation. On the~contrary, splitting a~radiological image into overlapping patches, or~generating new content via a~type of~Generative Adversarial Network are rarely used.

\begin{table}[!ht]
\caption{Data augmentation techniques used in~studies. Some techniques are parametrizable, so the~table indicates the~techniques and~parameters used. An indentation is used to show the~subtypes of~the~method.}
\label{tab:augmentation}
\centering
\begin{tabular}{|l|l|}
\hline
\textbf{Data augmentation technique}          & \textbf{Values and~studies}                     \\ \Xhline{2\arrayrulewidth}

Affine transformations & \cite{b13}: \\ \hline
\hspace{3mm} Rotation & \cite{ b27, b29, b12, b23}, 5\textdegree \cite{b6}, 15\textdegree \cite{b8, b16, b24}, 20\textdegree \cite{b11}, 25\textdegree \cite{b25} \\ \hline
\hspace{3mm} Scaling / Zooming & \cite{b29, b11, b23}, 10\% \cite{b6, b15, b25}, 20\% \cite{b27} \\ \hline
\hspace{3mm} Flip & \cite{b11, b15, b17} \\
\hspace{8mm} Horizontal & \cite{b29, b30, b6, b12, b23} \\
\hspace{8mm} Vertical & \\ \hline
\hspace{3mm} Shifting / Translation & \cite{b29, b11, b23}, height 5\% \cite{b6}, 10\% \cite{b25} \\ \hline
\hspace{3mm} Shearing & \cite{b27, b28, b11} \\ \Xhline{2\arrayrulewidth}
Brightness change & \cite{b13, b23, b29} +/-30 \cite{b28}, 10\% \cite{b6} \\ \hline
Crop & \cite{b30, b12} \\ \hline
Contrast change & \cite{b13} \\ \hline

Gaussian noise & \cite{b28} \\ \hline
ZCA whitening transformation & \cite{b11} \\ \hline

Elastic transformation & $\alpha$=60, $\sigma$=12 \cite{b25} \\ \hline
Grid distortion & steps=5, limit=0.3 \cite{b25} \\ \hline
Optical distortion & distort=0.2, shift=0.05 \cite{b25} \\ \hline
Warping & 10\% \cite{b15} \\ \Xhline{2\arrayrulewidth}

Multiple patches from~each image & \cite{b19} \\ \hline
Class-inherent transformations Network* & \cite{b26} \\ \Xhline{2\arrayrulewidth}

Augmentation used but parameters are not specified & \cite{b7} \\ \hline
No augmentation used & \cite{b18, b20, b21, b5, b9, b10, b31} \\ \hline

\end{tabular}

* inspired by~Generative Adversarial Networks
\end{table}

However, not all of~them are appropriate from~a~medical point of~view. Before an augmentation, it~is~recommended to consider the~'safety' for~the~chosen domain. For~example, the~rotation should be done carefully, because some parts of~the~lungs, such as costophrenic recesses, may be placed outside the~image. Also, change of~brightness or~contrast should be performed only in~a~limited manner, as greater manipulation may obscure lung structure. Moreover, in predicting COVID-19, it is acceptable to~crop or~proportionally scale/zoom an~image to such an~extent that it displays only the~lungs without a~background or~other parts of~the~body. 

It is also worth noting, that in~the~case of~CT and~X\=/ray images, the~augmentation based on~rotation or~flipping generate photos that cannot naturally appear in~real datasets, because the~process of~taking the~photo itself is standardized. Horizontal flips should be done carefully, with some specific limitations. Most pathologies will be present similarly on the left or~right lung, except for the change in shape of~the~heart (like in dextrocardia) or~pathologies affecting specific lobes, due to different lung anatomy (like lobar pneumonia or~lobar atelectasis). These limitations should be taken into consideration in model design. 

In~general, all augmentation methods should be consulted with radiologists, as domain knowledge is~crucial. In~every project, it is important to know the~field of~research to avoid a~situation in~which instead of~solving the~problem, bias is accidentally introduced.

\subsection{Model architecture}

In~the~studies different approaches of~modeling were applied. Some benefited from~machine learning methods, whereas the~rest used deep learning. In~the~first case, simple classifiers or~their ensembles were applied: AdaBoost \cite{b7}, KNN \cite{b7}, Naive Bayes \cite{b7}, SVM \cite{b7}.

In~the~reviewed studies, lung-specific model architectures (own models) were relatively often used for~classification, whereas the existing architectures were frequently fine-tuned. The~following model architectures or~their fine-tuned, modified versions were investigated: ResNet \cite{b16} (ResNet18 \cite{b19, b9, b17, b25}, ResNet34 \cite{b20, b9, b15}, ResNet50 \cite{b18, b5, b6, b12, b15}), DenseNet \cite{b16, b10, b23, b25} (DenseNet121 \cite{b6, b24}, DenseNet-161 \cite{b9}, DenseNet-201 \cite{b5, b15, b24}), VGG \cite{b16, b25} (VGG-16 \cite{b8, b5, b24, b31},VGG-19 \cite{b5, b15, b24}), Inception \cite{b25}, InceptionV3 \cite{b27, b9}, InceptionResNetV2 \cite{b5, b9, b24}, MobileNetV2 \cite{b5, b24}, NASNetMobile \cite{b5, b24}, EfficientNet-B0 \cite{b15}, Efficient TBCNN \cite{b6}, MobileNet \cite{b24}, NASNetLarge \cite{b24}, Res2Net \cite{b30}, Attention-56 \cite{b24}, ResNet15V2 \cite{b5}, ResNet50V2 \cite{b11}, ResNeXt \cite{b9}, WideResNet \cite{b9}, Xception \cite{b24}, own model \cite{b13, b21, b7, b28, b29, b15, b26, b25}. It is clearly visible that there are numerous types of~neural networks. Different neural networks can catch different dependencies in~the data. For~solving a~problem, many types of~model architectures are tested to find the~best one for~a~specific task. Recommendations on~how the~explanations should look do not depend on~the~neural network architecture.

For segmentation, the~following architectures were used: U-net \cite{b26, b12, b25}, AutoEncoder \cite{b17}, VGG-16 backbone + enhanced feature module \cite{b30}, (FC)-DenseNet-103 \cite{b19}, Nested version of~Unet (Unet++) \cite{b25}, VB-Net \cite{b20}. During the~segmentation process, it is important that the~lungs are accurately segmented. Otherwise, distorted border lines can be an indication of~pathology. In~study \cite{b19}, the~authors were aware that their segmentation cut pathological changes in~lungs. In~study \cite{b25} segmentation for~non-domain experts appears accurate. However, radiologists noticed that also other structures (i.e., bowel loops) were interpreted as lungs in~that segmentation.

There are multiple purposes for~creating new model architectures. The~most common is adjusting existing architectures for~better explainability or~scalability for~training on~medical COVID\=/19 imaging \cite{b13, b7}. For~example, in~studies \cite{b29, b15}, the~authors conducted tests and~chose the~advantages of~many architectures while creating their own. The~proposed architectures are usually smaller and~require a~lower number of~trainable parameters than in~well-known DNN architectures \cite{b26, b28}.

Six studies published their code on~GitHub: \cite{b19, b27, b29, b6, b10, b23}. Other studies did not include any reference to their code or~model. 

Often the~prediction from~multiple models is combined to improve the~overall performance. However, surprisingly, in~the~reviewed studies, there were not many ensemble models: \cite{b16, b20, b9, b25}.

\subsection{Transfer learning}

Transfer learning is an~ML technique about reusing gained knowledge from~one problem to a~new one. In~the~reviewed studies, it is commonly used when the~neural network has a large number of~parameters or~the~number of~collected samples is too small for~a~specific task. In~such~a~case, fewer training epochs are needed to adjust the~model to a~particular task. There are several popular image databases: ImageNet and, NoisyStudent for~which various architectures of~pre-trained neural networks are available. Transfer learning on~ImageNet database was utilized in~the following studies: \cite{b8,b19,b26,b27,b28,b29,b5,b6,b7,b15,b24,b31}. Twelve out of~25 studies decided to use a~neural network pre-trained on~ImageNet for~transfer learning. Therefore, it can be said that this is a~very common procedure.

However, as \cite{Cheplygina2018} shows, it is not clear whether using ImageNet for~transfer learning in~medical imaging is the~best strategy. ImageNet consists of~natural images. Meanwhile, medicine is an~entirely different field and~is completely unrelated. \cite{Kim2017} also stressed the~fact that the~features which are extracted by~models pre-trained on~ImageNet can introduce bias.

Only in~three of~the~reviewed studies was transfer learning conducted on~lung images. The chosen datasets included 112,120 in~\cite{b23}, 88,079 in~\cite{b10}, and~951 in~\cite{b25} non-COVID\=/19 lung images. The~study \cite{b16} did not perform any transfer learning because lung images lack colorful patterns, specific geometrical forms, or~similar shapes. The amount of~redundant information introduced by a network pre-trained with color images may seriously affect the learning process on gray level images. In~study \cite{b6}, the~authors discovered that the~model has better performance when pre-trained on~ImageNet than without it. However, the~authors found out that their models pre-trained on~ImageNet were using irrelevant markers on~lung images while making a~prediction.

Especially when the~model is trained on~a~small amount of~data, the~usage of~completely irrelevant features from~another pre-trained model may increase model accuracy/result. For~this reason, it is crucial to find a~large database with images similar in~domain and~appearance to limit the~possibility of~irrelevant markers that take part in~a~prediction. It is recommended to train a~neural network on~this database and~then use transfer learning to adjust it to the~target task.

For transfer learning, it is recommended to take into consideration the~following X\=/ray data sources with DICOM images (consider the~fact that, in~some of~them, children and~adults lungs are mixed): U.S. National Library of~Medicine\footnote{pubmed.ncbi.nlm.nih.gov/25525580} ($7,470$ images), Radiological Society of~North America\footnote{kaggle.com/c/rsna-pneumonia-detection-challenge} ($29,684$ images), Society for~Imaging Informatics in~Medicine\footnote{kaggle.com/c/siim-acr-pneumothorax-segmentation} ($3,209$ images), Medical Information Mart for~Intensive Care\footnote{physionet.org/content/mimic-cxr/2.0.0} ($377,110$ images).
For transfer learning on~CT, the~following data sources are available:
The Reference Image Database to Evaluate Therapy Response\footnote{wiki.cancerimagingarchive.net/display/Public/RIDER+Lung+CT} ($15,419$ images), A Large-Scale CT and~PET/CT Dataset for~Lung Cancer Diagnosis\footnote{wiki.cancerimagingarchive.net/pages/viewpage.action?pageId=70224216} ($260,826$ images), The~National Lung Screening Trial\footnote{wiki.cancerimagingarchive.net/display/NLST} ($21,082,502$ images).

\subsection{Training parameters}

The selection of~hyperparameters has a~large impact on~model results. Nevertheless, the~process of~tuning parameters is empirical and~depends on~the~model architecture. For~this reason, it is difficult to present a~set of~parameters adequate for~every model architecture. However, there are several tips which can be used for~most models. 

Often the~learning rate is decreased during the~training process. Sometimes callback functions are used to halt training, when the~result of~a~model is optimum, and~during the~training process, to save and~store the~best model and~its~parameters. The~most typically used optimizer is Adam. The~batch size of~images during model training is between 2 and~81 with the~most common value 8.

The whole image dataset is typically divided into 3 or~2 sets, most commonly into: training set 80\%, validation set 10\% and~testing set 10\% \cite{b26, b28, b23, b31}. Proportion 80\% to~20\% was the~most frequently used among divisions into training and~testing set respectively \cite{b5, b11}.

In~study \cite{b5}, the~recommendation to conduct external validation is indicated, meaning an~evaluation on~an~independent database. Another public dataset will be the~best choice for~cross-database validation \cite{b19}. However, in~the~reviewed studies cross-validation is the~most frequently used. It is a~common choice for~training on~a~small amount of~data resources. The~problem which may occur during cross-validation is~overfitting to the~data. For~this reason, validation on~an~external resource is the~most trustworthy method.

\subsection{Model performance}

Evaluation metrics are commonly used to compare different models. For~DNN image classification, there are many metrics frequently used for~model quality assessment. In~the~reviewed studies, we discovered a~large variability in~the number of~reported metrics. It is a~common situation due to the~fact that there are no detailed recommendations as to which performance metric should be used. We recommend the~instructions presented in~\cite{Albahri2020}, but, unfortunately, in~almost all the~reviewed studies, at~least one metric out of~these recommendations was missing.

Based on~the~rules described in~study \cite{Albahri2020}, there are six evaluation criteria for~binary classification: accuracy, precision, recall (sensitivity), F score, specificity, AUC. For~multi-class classification, there are eight criteria: average accuracy, error rate, precision$_\mu$, recall$_\mu$, F score$_\mu$, precision$_M$, Recall$_M$, F score$_M$, and~for multi-label classification four criteria: exact match ratio, labelling F score, retrieval F score, Hamming loss. 

There is another important factor which indicates why more than one evaluation metric should be used. It provides the~opportunity to compare model architectures and~then choose the~best one for~a~given problem. Nevertheless, the~models were not trained on~the~same images. Some databases contained only severe cases which were easier to classify \cite{b26}. Even if~studies refer to the~same data resources, it is possible that the~amount of~data has increased over time. For~this reason, it is rather difficult to make a~reliable comparison. The~most trustworthy way to compare different model architectures is to look at studies which tested many of~them, i.e. \cite{b26, b5, b7, b15, b24, b25}.

\section{Explainable artificial intelligence}

\subsection{The importance of model explainability}

When designing predictive models for healthcare, or~any other high-stakes decisions, the explainability of~the model is a key part of~the solution. The empirical performance of~the model is very important, but there can be no responsible modeling if the issue of~explainability is not addressed properly for each stakeholder of~the system.
For physicians, the lack of~explainability drastically reduces the confidence in the system. For model developers, it makes it difficult to detect flaws in model behavior and obstructs debugging \cite{biecekburzykowski, Holzinger2017}.

\begin{figure}[!ht]
\centering
\begin{tabular}{p{0.3\textwidth} p{0.3\textwidth} p{0.3\textwidth}}
  a) \includegraphics[width=.3\textwidth]{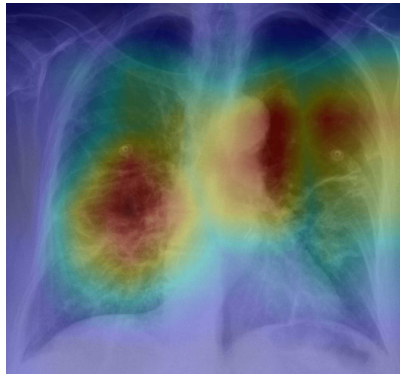}\hfill &
  b) \includegraphics[width=.3\textwidth]{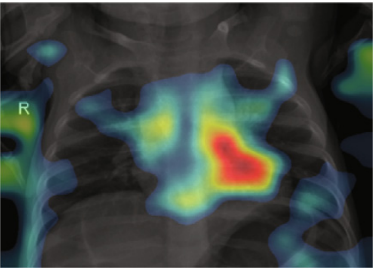}\hfill &
  c) \includegraphics[width=.3\textwidth]{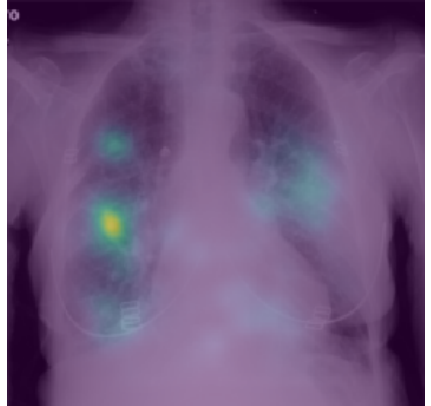}\hfill \\
  d) \includegraphics[width=.3\textwidth]{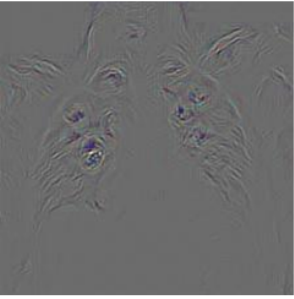}\hfill &
  e) \includegraphics[width=.3\textwidth]{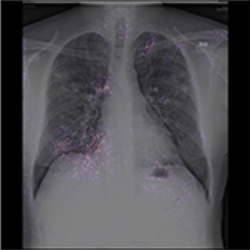}\hfill &
  f) \includegraphics[width=.3\textwidth]{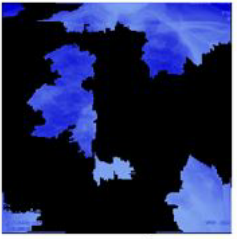}\hfill
\end{tabular}
\caption{Examples of~explanations for~COVID-related models from~studies: \cite{b26, b28, b5, b9, b10, b11}. The~following explanations are used: a) Grad-CAM, b) CAM, c) saliency, d) guided backpropagation, e) integrated gradients, f) LIME. Such explanations can be divided into 4 types: heat maps (image a) - c)), contour lines (d)), points (e)), and~image pieces (f)).}
\label{fig:xaiExamples}
\end{figure}

For predictive models, two general approaches to explainability are either by using classes of~interpretable-by-design models or~using post-hoc explanations. Despite the obvious advantages of~interpretable-by-design models, their construction requires more domain knowledge linked to the construction of~interpretable features. The advantage of~post-hoc explanations is that they are constructed after the model has been trained. Thus, the developer can focus on model performance by pouring large volumes of~data into a~neural network and then deal with model explanations afterwards. In the analyzed studies, the authors used only post-hoc methods which are prevalent in computer vision tasks \cite{Samek2017}. Examples of~post-hoc explanations are presented in Figure \ref{fig:xaiExamples}.

Due to the mode of~operation, also post-hoc explanation methods can be divided into two groups. The~first group consists of~input perturbation methods such as Locally Interpretable Model Explanations (LIME), or~Occlusion Sensitivity. These methods are based on the analysis of~the change of~the model response after obscuring, removing, or~perturbing some part of~the image. The advantage of~this class of~methods is that they are insensitive to the internal structure of~the model. Such so-called model-agnostic approaches assume nothing about the internal structure of~the model. By analyzing how a series of~input perturbations affect the final prediction, it determines which part of~the input is important.

The second group are methods based on the analysis of~signal propagation through the network, i.e. model-specific methods. This group of~methods uses detailed information about the network architecture and the design of~subsequent layers to determine the key regions of~input for the final prediction. The~advantage of~such approaches is that usually, one pass through the structure of~the network is sufficient to generate explanations. Model specific methods for explanations of~CNNs can be organized into a spectrum of~solutions, from gradient-based methods to activation map-based methods.

For gradient-based methods, the gradient $\frac{dy}{dx}$ between the output model class $y$ and the input image $x$ is used to calculate saliency maps. For large networks, such as most of~those shown in Table \ref{tab:neural_networks}, the gradient information is very noisy, so there have been many modifications to this method that reduce noise by smoothing or~thresholding or~rescaling. This~class of~models includes Guided Backpropagation, Layer-wise relevance propagation, and SmoothGrad.

\begin{table}[!ht]
\caption{The depth, number of~parameters and type of~layers for neural networks in considered papers. For large networks gradient based explanations are noisy. Some explanation techniques assume specific types of~layers.}
\label{tab:neural_networks}
\centering

\begin{tabular}{|p{0.18\textwidth}|p{0.06\textwidth}|p{0.08\textwidth}|p{0.58\textwidth}|}
\hline
\textbf{Model architectures}   & \textbf{Depth} & \textbf{No. of parameters} & \textbf{Layer types} \\ \hline
ResNet18, ResNet34, ResNet50, ResNet15V2, ResNet50V2            & - & 11.7M-25.6M & ZeroPadding2D, Conv2D, BatchNormalization, Activation, MaxPooling2D, Add, GlobalAveragePooling2D, Dense \\ \hline 
DenseNet121, DenseNet-161, DenseNet-201  & 121-201 & 8.1M-20.0M & ZeroPadding2D, Conv2D, BatchNormalization, Activation, MaxPooling2D, Concatenate, AveragePooling2D, GlobalAveragePooling2D, Dense \\ \hline 
VGG-16, VGG-19            & 23-26 & 138-144 & Conv2D, Dense, Flatten, InputLayer, MaxPooling2D\\ \hline 
InceptionV3                    & 159 & 23.9M & Conv2D, BatchNormalization, Activation, MaxPooling2D, AveragePooling2D, Concatenate, GlobalAveragePooling2D, Dense\\ \hline 
InceptionResNetV2              & 572 & 55.9M & Conv2D, BatchNormalization, Activation, MaxPooling2D, AveragePooling2D, Concatenate, Lambda, GlobalAveragePooling2D, Dense\\ \hline 
MobileNet                      & 88 & 4.3M & Conv2D, BatchNormalization, ReLU, DepthwiseConv2D, ZeroPadding2D, GlobalAveragePooling2D, Reshape, Dropout, Activation \\ \hline
MobileNetV2                    & 88 & 3.5M & Conv2D, BatchNormalization, ReLU, DepthwiseConv2D, ZeroPadding2D, Add, GlobalAveragePooling2D, Dense\\ \hline 
NASNetMobile, NASNetLarge                   & - & 5.3M-88.9M & Conv2D, BatchNormalization, Activation, ZeroPadding2D, SeparableConv2D, Add, MaxPooling2D, AveragePooling2D, Cropping2D, Concatenate, GlobalAveragePooling2D, Dense\\ \hline
EfficientNet-B0                & - & 5.3M & Rescaling, Normalization, ZeroPadding2D, Conv2D, BatchNormalization, Activation, DepthwiseConv2D, GlobalAveragePooling2D, Reshape, Multiply, Dropout, Add, Dense\\ \hline 
Efficient TBCNN                &  & 0.23M  & Conv2D, MaxPool2D, BatchNormalization, GlobalAveragePooling2D, Add, Dense \\ \hline 
Attention-56                   & 115 & 31.9M &Conv2D, Lambda, MaxPool2D, UpSampling2D, AveragePooling2D, ZeroPadding2D, Dense, Add, Multiply, BatchNormalization, Dropout \\ \hline
Xception                       & 126 & 22.9M & Conv2D, BatchNormalization, Activation, SeparableConv2D, MaxPooling2D, Add, GlobalAveragePooling2D, Dense \\ \hline
\end{tabular}

\end{table}

Methods based on activation maps, such as Class Activation Mapping (CAM) or~DeepLIFT, focus on visualizing the relationship between the layer with the feature map (in most cases the penultimate layer of~the~network) and the model output. Assuming that the feature map stores information about the spatial relevance of~features, one can explore what elements of~the feature map are most relevant for the final prediction. Such methods often have an assumption about the structure of~the network, such as global average pooling before the softmax layer.

In our analysis, the most popular solution turned out to be the one that combines both mentioned above approaches, tracing the gradient between the model prediction and the feature map and then analyzing the~spatial information of~a specific part of~the feature map. This group of~methods includes the most popular explanation method Grad-CAM and its modifications Guided Backpropagation, Guided Grad-CAM, Grad-CAM++. Using gradient tracking between the feature map and network output is also a more flexible approach in terms of~network architecture without enforcing global pooling.

\subsection{XAI methods used in~the~reviewed studies}

The area of~model explanations and~the number of~methods that can be used for~this purpose are increasing rapidly \cite{Holzinger2017}. Such methods differ in~properties; they work either for~a~single image (so-called instance level methods) or~globally for~the~whole dataset. Some of~them are based on~gradients, others on~interpretable features, some are intrinsic or~post-hoc, model-specific (class-discriminative, high-resolution, multi-layer) or~model-agnostic. 

Table \ref{tab:xai} shows which approach to model explanation was used in~each study. The~most popular in~the~reviewed studies was Grad-CAM. Its popularity may be related to the~fact that colorful heat maps are easy to~implement and~seem to be readable. An example of~an implementation for~Grad-CAM is available online, and~its use on~melanoma images shows great results.

\begin{table}[!ht]
\caption{XAI techniques used in considered papers}
\label{tab:xai}
\centering
\begin{tabular}{|l|l|}
\hline
\textbf{Name of the XAI technique}                                & \textbf{Reference}                                  \\
\hline
Grad-CAM (gradient-weighted class activation mapping) & \cite{b8, b13, b16, b18, b19, b20, b21, b26, b27, b6, b23, b25} \\ \hline
LIME (local interpretable model-agnostic explanations) & \cite{b5, b15, b25}                           \\ \hline 
CAM (class activation mapping)                         & \cite{b20, b28, b5, b11}                           \\ \hline
Saliency (saliency map)                                & \cite{b6, b9, b10, b11}                            \\ \hline
Guided Backpropagation                                 & \cite{b9, b11}                                     \\ \hline
LRP (layer-wise relevance propagation)                 & \cite{b16, b26}                                    \\ \hline 
Occlusion (occlusion sensitivity)                      & \cite{b18, b9}                                     \\ \hline 
AM (activation mapping)                                & \cite{b30}                                         \\ \hline
Attribution maps                                       & \cite{b17}                                         \\ \hline
DeepLIFT                                               & \cite{b9}                                          \\ \hline 
Feature maps                                           & \cite{b24}                                         \\ \hline 
Grad-CAM++                                             & \cite{b16}                                         \\ \hline
Guided Grad-CAM                                        & \cite{b11}                                         \\ \hline
GSInquire                                              & \cite{b29}                                         \\ \hline
Input X Gradient                                       & \cite{b9}                                          \\ \hline 
Integrated Gradients                                   & \cite{b9}                                          \\ \hline 
\end{tabular}
\end{table}

Another very popular method is LIME. As clearly visible in~studies \cite{b5, b15}, some large superpixels include different structures (i.e., lung tissue and~chest wall). Therefore, this method is not accurate enough for an~interpretable representation of~space due to the lack of~semantic meaning.

Some threads related to the~application of~XAI in~the analyzed publications are questionable.
Contrary to what \cite{b16} states, explanations of~ensemble models are possible. A~single best model does not have to be selected for~the~visualization of~the~prediction. There are many model-agnostic interpretation methods which do not rely on model architecture and can be easily used for~explanations, such as: LIME, or~Anchors. In~most XAI methods, it is possible to adjust them to suit ensemble models. Moreover, ensemble models usually outperform a~single model in~terms of~accuracy.

According to \cite{b21}, the~model makes incorrect decisions in~poor quality X\=/rays. This is because the~low quality or~very low-resolution images do not show enough details even for~the~models. Such images should be removed while checking the~database contents.
    
The study \cite{b18} noticed that the~region of~the~lesion is marked correctly, but~that~model prediction is wrong. Unless we perform a~quantitative and~qualitative evaluation of~XAI results, we will not have the~opportunity to assess the~trustworthiness of~our model. The~model may take into greater consideration other image features than it should. To explore this kind of~a~model mistake, other XAI methods ought to be used to obtain a~better comparison possibility.

\subsection{Domain experts' evaluation of XAI methods}

In~most of~the~reviewed studies, the~application of~XAI comes down to~the~series of~colorful images without any assessment about how valid these explanations are. Colored explanations obscure the~original image, which makes it even more difficult to assess their correctness. In images with XAI heat maps, it is often hard or~impossible to see pathologies and~guess if~the~model works well. Raw lung images shall be put next to explanations.
Also the~explanations should be interpreted or~validated by~radiologists. Otherwise, they are redundant and~contribute nothing to the~trustworthiness of~the~model.

Together with the~radiologists, we analyzed the~explanations from~the~discussed works. In~the~following paragraphs, we discuss the~most common mistakes or~inappropriate explanations.

\begin{figure}[!ht]
\centering
\begin{tabular}{p{0.36\textwidth} p{0.36\textwidth}}
  a) \includegraphics[width=.3\textwidth]{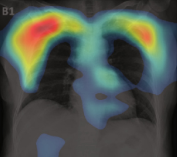}\hfill &
  b) \includegraphics[width=.3\textwidth]{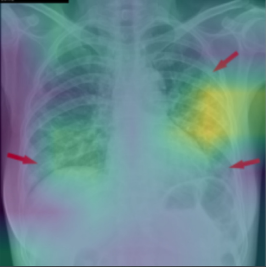}\hfill \\
  c) \includegraphics[width=.3\textwidth]{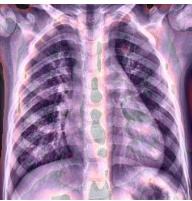}\hfill &
  d) \includegraphics[width=.3\textwidth]{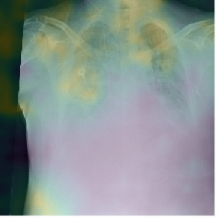}\hfill
\end{tabular}
\caption{Examples of~biased model explanations: a) \cite{b28}, b) \cite{b8}, c) \cite{b6}, d) \cite{b17}. Red arrows in~the image b) are marked by~a~radiologist to help locate the~lesions. They were not present in~the~training set.}
\label{fig:wrongModel}
\end{figure}

In~the~first example, in~Figure \ref{fig:wrongModel}a), the~model focuses on~clavicles, scapulas, and~soft tissues, which are outside the~lungs. Very likely, the~model predicts illness based on~an~improper part of~the~image. Location of~the~areas marked by~explanation should be inside the~chest on~the~lung tissue because COVID\=/19 lesions are not located on, e.g., lymph nodes. Moreover, there are some elements that cannot be considered as decision factors like imaging artifacts (cables, breathing tubes, image compression) or~embedded markup symbols \cite{b29}. To prevent the~model from~focusing on~irrelevant features, in~some studies, the~lungs were segmented, and~their background was removed \cite{b19, b20, b26, b12, b25}. However, it may not help when some imaging artifacts are present in~the area of~the~lungs.

The second example, in~Figure \ref{fig:wrongModel}b) shows that the~model does not take the~lesions into account. The~model states that parts of~the~lungs other than the~ones marked by~the~radiologist are relevant for~model prediction. Explanations that “roughly indicate the~infection location” \cite{b20} are not acceptable for~the~robust model. The~model should do this with the~accuracy of~the~pixel marked by~radiologists as relevant.

The third example, in~Figure \ref{fig:wrongModel}c), visualization is not clear. The~study describes a~different XAI method than the one present in~the image. Moreover, this visualization highlights the~whole image, and~it is not possible to guess which features took part in~the prediction. It is important to point out that some explanation methods can give clearer results for~a~specific type of~DNN and~for~a~specific domain.

The last example, in~Figure \ref{fig:wrongModel}d) is blurred. The~image of~the~lungs is improperly taken, and~the process should be repeated. The~current image is useless for~the~accurate diagnosis process. Such images should be removed during data resource verification before model training.

If the~lung lesions are well described, it will be possible to prepare quantity and~quality XAI assessment to score the~trustworthiness of~the~specific model. One possible option would be to create measures for~the~evaluation of~XAI image models based on~the~measures quoted in~study \cite{DeYoung2019}: Intersection-Over-Union and~token-level, which presents measures for~the~evaluation of~text models.

Evaluation of~explanation methods is crucial for~confirmation of~model trustworthiness. First of~all, radiologists should validate a~specific model with the~help of~XAI. They should assess location, size, and~shape of~marked regions by explanation methods. Their interpretations should contain clear references to structures and~lesions in~the~lungs, such as posterior basal segment, ground-glass opacity, consolidation, frosted glass shadows, etc. 
The example of~a~well-prepared XAI interpretation can be found in~the~study \cite{b18}.

\section{The checklist for~responsible analysis of~lung images with deep learning models}

In~this work, we have shown that development of~a~model which analyzes lung images is a~complex process. Therefore, we prepared the checklist based on the~analyzed studies and~the~errors we found in~them. In~\cite{gawande2011checklist}, it is shown that well-prepared checklists significantly improve the~quality of~the~modeling process. They help to avoid, or~quickly detect and~fix, errors.

In~the~list below, the~letter R indicates that the~point should be consulted with a~field expert / radiologist, and~the~letter D indicates that the~point should be consulted with a~model developer.

The points in~the~checklist below are grouped according to the~sections' names discussed in~this study. This should assist in~finding a~detailed description of~the~problem stated in~the~checkpoint list in~the~corresponding section.

\begin{itemize}
\item Data resources
    \begin{todolist}
    \item [D] Does the~data and its associated information provide sufficient diagnostic quality? If~images are in~DICOM, does the header provide the needed information? If~not, is it provided in~any other way?
    \item [R] Are the~low quality images (i.e., blurred, too dark, or~too bright) rejected?
    \item [D] Is the~dataset balanced in~terms of~sex and~age?
    \item [R] Does the~dataset contain one type of~images (CT or~X\=/ray)? 
    \item [R] Are the~lung structures visible (“lung” window) on~CT images?
    \item [D] Are images of children and of adults labelled as such within the dataset?
    \item [R] Are images correctly categorized in~relation to class of~pathology?
    \item [D] Are AP/PA projections described for~every X\=/ray image?
\end{todolist}
\item Image preprocessing
    \begin{todolist}
    \item [D] Is the~data preprocessing described?
    \item [D] Are artifacts (such as captions) removed?
    \end{todolist}
\item Data augmentation (if needed)
    \begin{todolist}
    \item [D] Are the~lungs fully present after transformations?
    \item [R] Are lung structures visible after brightness or~contrast transformations?
    \item [D] Are only sensible transformations applied?
\end{todolist}
\item Transfer learning (if used)
    \begin{todolist}
    \item [D] Is the~transfer learning procedure described?
    \item [D]  Is the applied transfer learning appropriate for this case (i.e.: images of same type and content have been used to train the original model)?
\end{todolist}
\item Model performance 
    \begin{todolist}
    \item [D] Are at least a~few metrics of~those~proposed in~\cite{Albahri2020} used?
    \item [D] Is the~model validated on~a~different database than the~one used for~training?
\end{todolist}   
\item Domain quality of~model explanations
    \begin{todolist}
    \item [R] Are other structures (i.e., bowel loops) misinterpreted as lungs in segmentation?
    \item [R] All the areas marked as~highly explanatory are located inside the~lungs?
    \item [R] Are artifacts (cables, breathing tubes, image compression, embedded markup symbols) misidentified as part of~the~explanations?
    \item [R] Are areas indicated as explanations consistent with opinions of~radiologists?
    \item [R] Do explanations accurately indicate lesions?
     \end{todolist}
\end{itemize}

\begin{table}[!ht]
\caption{Summary showing which points from the checklist are fulfilled by the reviewed data resources.}
\label{tab:summary_of_data_resources}
\scalebox{0.85}{
\begin{tabular}{|p{0.3\textwidth}|p{0.03\textwidth}|p{0.03\textwidth}|p{0.03\textwidth}|p{0.03\textwidth}|p{0.03\textwidth}|p{0.03\textwidth}|p{0.03\textwidth}|p{0.03\textwidth}|p{0.03\textwidth}|p{0.03\textwidth}|p{0.03\textwidth}|p{0.03\textwidth}|p{0.03\textwidth}|p{0.03\textwidth}|p{0.03\textwidth}|}

\hline

\textbf{Checklist / Data resource} 	&2)  	&3)  	&4)  	&5)  	&6)  	&7)  	&8)  	&9)  	&10) 	&11) 	&12) 	&13) 	&14) 	&17)	&23) 	\\ \hline
[D] Does the~data and its associated information provide sufficient diagnostic quality? & Y? & N? & N & N? & N? & N & N? & N & N? & N & N & N & Y & N? & N \\ \hline
[R] Are the~low quality images rejected? 	& N 	& N 	& N 	& N 	& N 	& N 	& N 	& N & N 	& n/a 	& Y 	& N 	&	N	& Y?	& N 	\\ \hline
[D] Is the~dataset balanced in~terms of~sex and~age? & Y? & ? & ? &Y? & Y? & Y & Y& ? & N & N & ? &? &? &Y?&? \\ \hline
[R] Does the~dataset contain one type of~images (CT or~X\=/ray or the same projection)? 	& Y 	& Y 	& Y 	& N  	& Y 	& Y	& N &N	& Y 	& Y 	& Y 	& Y 	&	N & Y  & N 	\\ \hline
[R] Are the~lung structures visible (“lung” window) on~CT images? 	&  n/a 	& n/a 	& n/a 	& n/a 	& n/a 	& n/a	& n/a  &	n/a& n/a 	& N 	& n/a 	& n/a 	&	n/a	& n/a  & n/a 	\\ \hline
[D] Are images of children and of adults labelled as such within the dataset? & not all & N & N & Y?& Y & Y & Y & N & Y & not all &N & N&N &N & N \\ \hline
[R] Are images correctly categorized in~relation to class of~pathology?	& N	& N	& Y	& N	& N	& N & N & N	& Y	& N	& Y	& N	&	Y	& N? & N	\\ \hline
[D] Are AP/PA projections described for~every X\=/ray image?&N&Y&N&Y&Y&Y&Y&N& Y&n/a&N&N&Y&Y&N \\ \hline

\end{tabular}
}
\end{table}

\begin{table}[!ht]
\caption{Summary showing which points from the checklist are fulfilled by the peer-reviewed studies. }
\label{tab:summary}
\centering
\newcolumntype{L}[1]{>{\raggedright\let\newline\\\arraybackslash\hspace{0pt}}m{#1}}
\def\arraystretch{1.4}
\setlength{\tabcolsep}{2.4pt}

\begin{tabular}{|p{0.45\textwidth}|l|l|l|l|l|l|l|l|l|l|l|l|}
\hline
\textbf{Checklist / Study} & \textbf{\cite{b8}} &  \textbf{\cite{b13}} &  \textbf{\cite{b16}} &  \textbf{\cite{b18}} & \textbf{\cite{b19}} & \textbf{\cite{b20}} & \textbf{\cite{b21}} &  \textbf{\cite{b26}} & \textbf{\cite{b27}} & \textbf{\cite{b28}} & \textbf{\cite{b29}} & \textbf{\cite{b30}} \\ \hline

 \multicolumn{13}{|l|}{\textbf{Image preprocessing}} \\ \hline
[D] Is the~data preprocessing described? & Y & N & Y & Y & Y & Y & Y & Y & Y & Y & Y & Y \\ \hline 
[D] Are artifacts (such as captions) removed? & ? & ? & Y & Y & Y & n/a & ? & Y & N & N & Y & n/a \\ \hline
 \multicolumn{13}{|l|}{\textbf{Data augmentation (if needed)}} \\ \hline
[D] Are the~lungs fully present after transformations? & ? & ? & ? & n/a & Y? & Y & n/a & ? & ? & ? & ? & N?\\ \hline 
[R] Are lung structures visible after brightness or~contrast transformations? & n/a & ? & n/a & n/a & n/a & Y & n/a & ? & n/a & ? & ? & n/a \\ \hline 
[D] Are only sensible transformations applied? & Y & ? & Y & n/a & Y & Y & n/a & ? & ? & N & N & N \\ \hline
 \multicolumn{13}{|l|}{\textbf{Transfer learning (if used)}} \\ \hline
[D] Is the~transfer learning procedure described? & Y & n/a & n/a & Y? & Y & n/a & Y & Y & Y & Y & Y & Y \\ \hline 
[D] Is the applied transfer learning appropriate for this case? & N & n/a & n/a & N & N & n/a & N & Y?  & N & N & N & N \\ \hline
 \multicolumn{13}{|l|}{\textbf{Model performance}} \\ \hline
[D] Are at least a~few metrics of~those~proposed in~\cite{Albahri2020} used? & Y & Y & Y & Y & Y & Y & Y & Y & Y &Y & Y & Y\\ \hline
[D] Is the~model validated on~a~different database than the~one used for~training? & N & N & N & N & Y & Y & N & N & N & N & N & N\\ \hline
 \multicolumn{13}{|l|}{\textbf{Domain quality of~model explanations}} \\ \hline
[R] Are other structures (i.e., bowel loops) misinterpreted as lungs in segmentation?& n/a & n/a&  n/a & n/a & N & Y? & n/a & N & n/a &n/a & n/a & Y \\ \hline
[R] All the areas marked as~highly explanatory are located inside the~lungs? & Y & n/a & Y & n/a & Y & Y & & Y? & Y & N & Y & Y  \\ \hline
[R] Are artifacts misidentified as part of~the~explanations? & Y & n/a & & N & n/a & n/a & n/a & n/a & N & n/a & n/a & n/a\\ \hline
[R] Are areas indicated as explanations consistent with opinions of~radiologists? & N & n/a & n/a & n/a & n/a & n/a & n/a & n/a & Y & n/a & n/a & n/a \\ \hline
[R] Do explanations accurately indicate lesions? & Y? & n/a & Y? & n/a & Y? & Y & N & N & N & Y? & N & Y \\ \hline 

\end{tabular}

\end{table}

According to the prepared checklist, in Table \ref{tab:summary_of_data_resources} and \ref{tab:summary}, we tried to analyze which points are fulfilled by~the~reviewed studies and the datasets used in these papers for the neural network training. Due~to~the~possibility of~changes in preprints, we only examined papers already published in journals.

We applied the following denotements: “\textit{Y}”, “\textit{N}” mean yes and no respectively (if an answer is probable then the additional “\textit{?}” is added), “\textit{?}” means there is no information provided, “\textit{n/a}” signifies that the~issue does not apply to a particular publication. Due to the fact that we can only evaluate the information contained in the article, the answers given to some questions from the checklist need to be clarified. 

The question regarding balance in the dataset (Table \ref{tab:summary_of_data_resources}) has two components. Sometimes, the dataset is balanced concerning only one criterion but not in terms of~the second one. In such cases, we put “\textit{Y?}” sign. We did similarly in cases where many metadata are missing, but there is a balance considering the existing data.

We would like to stress that mixed projections of~X-rays are present in databases: 5), 8), 23), both PA and lateral. Databases 5) and 9) contain also CT exam images. Furthermore, dataset 23) includes CT scanogram described as X-ray which is inappropriate for medical use. It is important to note that if~the~datasets in Table \ref{tab:summary_of_data_resources} are labeled as containing images incorrectly categorized by pathology class, then in~most cases patients should be classified into more than one pathology class. Moreover, in other databases, X-rays marked as pneumonia or~other disorders have no radiological findings.

In Table \ref{tab:summary}, for the papers~\cite{b27, b28}, the assessment of the presence of artifacts was made based on~the~images provided by the authors. In the question regarding the~visibility of~lungs after data augmentation, we put the value “\textit{N?}” because there were no details about applied random cropping. Such an answer reflects the fact that cropping can be risky especially if~the~parameters are not carefully chosen. In case of~\cite{b19}, we put “\textit{Y?}” as the authors honestly pointed out that their model for segmentation sometimes improperly generates masks when there is severe opacity. In such situations, the mask covers lungs only partially.

We would like to explain the criteria for assessing data augmentation and transfer learning. In the case of~the former, we put “\textit{N}” when horizontal flip was applied. In the latter, the main criterion was whether the~authors used a model pre-trained on ImageNet dataset. Such behavior is not recommended as natural scene images are significantly different from medical images. The biggest difference is the fact that X-ray and CT images are in grayscale unlike images in~ImageNet. In the reviewed papers, there were two approaches to~transfer learning: the backbone weights were frozen and the case where the pre-trained weights were just an initialization and were later trained. We~decided to put “\textit{N}” in both scenarios as both require artificial conversion of~grayscale images to RGB.

In Table \ref{tab:summary}, the difficulty to assess the solutions applied in the papers occurred also in the section requiring radiological expert knowledge. In the group “Domain quality of~model explanations”, we were not able to~check which of~these points were fulfilled by each of~the studies, as most publications contain only few or~even none images, which could be radiologically evaluated. Nevertheless, although our observations were made on~such~limited data, we were able to identify many mistakes in the radiologic background.

\section{Conclusion}

The sudden outbreak of~the COVID\=/19 pandemic has shown us how we need effective tools to support the~physicians. Deep neural networks can offer much in the analysis of~lung images, but responsible modeling requires very thorough model validation. Models without explanations create validation debt and~explanations without consultations with a radiologist are just an illusion of~validation.

This work shows a critical analysis of~25 state-of-the-art articles that use deep learning models based on~lung images to identify COVID\=/19. We have selected the most advanced papers in which the authors made extra effort to supplement the models with explanations. But even in this sample, it turns out that only in 7 out of~the 25 reviewed studies, the models were consulted with radiologists and regarding the~model explanations, they were validated only in three studies \cite{b21, b27, b25}.

 It is important to point out that, for medical examination, the most valuable are large-resolution images, especially in DICOM format. Surprisingly, out of~the considered datasets, DICOM COVID\=/19 cases are available only in one online dataset, and there are only four datasets with DICOM images for other lung diseases. The motivations for explaining models are commendable. Nevertheless, in many works, interpretations of~explanations and summaries are missing. The XAI method is not a conclusion in itself.
The~fact that the model provides correct explanations for a few images does not yet show that the model works properly. It would be good to quantitatively validate XAI methods. For this purpose, the help of~clinicians or~proper annotations prepared by radiologists beforehand are necessary.

The paper mentions a long list of~problems in modeling, but this analysis is not intended to criticize any of~the mentioned articles, these are state-of-the-art papers often published in prestigious journals. However, the analysis of~these articles makes one look critically at standards in AI for healthcare or~rather the lack of~them. 
We hope that this paper will initiate the process of~development of~standards for responsible AI solutions in healthcare. In this paper, we showed that the verification of~the XAI solutions for medical images is not only important but it is a must. 

Following the guidelines proposed in the paper, we created an online GitHub repository\footnote{https://github.com/Hryniewska/checklist} which can be maintained by the community working on AI models for image analysis in healthcare. This repository is intended to be a starting point for further development of~the proposed checklist to meet the evolving challenges in responsible modeling. We believe that if the proposed checklist is taken into account when building models, we will get better models.

\section{Acknowledgment}

We would like to thank the anonymous reviewers whose detailed comments and valuable suggestions greatly improved this paper. Work on this paper was funded by the IDUB against COVID\=/19 initiative at~the~Warsaw University of~Technology.

\bibliographystyle{unsrt}  
\bibliography{references} 

\begin{thebibliography}{10}

\bibitem{Xie}
Xingzhi Xie, Zheng Zhong, Wei Zhao, Chao Zheng, Fei Wang, and Jun Liu.
\newblock {Chest CT for Typical Coronavirus Disease 2019 (COVID-19) Pneumonia:
  Relationship to Negative RT-PCR Testing}.
\newblock {\em Radiology}, 296(2):E41--E45, 2020.

\bibitem{Fang_Yicheng}
Yicheng Fang, Huangqi Zhang, Jicheng Xie, Minjie Lin, Lingjun Ying, Peipei
  Pang, and Wenbin Ji.
\newblock {Sensitivity of Chest CT for COVID-19: Comparison to RT-PCR}.
\newblock {\em Radiology}, 296(2):E115--E117, 2020.
\newblock PMID: 32073353.

\bibitem{Wong_Ho}
Ho~Yuen~Frank Wong, Hiu Yin~Sonia Lam, Ambrose Ho-Tung Fong, Siu~Ting Leung,
  Thomas Wing-Yan Chin, Christine Shing~Yen Lo, Macy Mei-Sze Lui, Jonan
  Chun~Yin Lee, Keith Wan-Hang Chiu, Tom Wai-Hin Chung, Elaine Yuen~Phin Lee,
  Eric Yuk~Fai Wan, Ivan Fan~Ngai Hung, Tina Poy~Wing Lam, Michael~D. Kuo, and
  Ming-Yen Ng.
\newblock {Frequency and Distribution of Chest Radiographic Findings in
  Patients Positive for COVID-19}.
\newblock {\em Radiology}, 296(2):E72--E78, 2020.
\newblock PMID: 32216717.

\bibitem{Corman}
Victor~M. Corman, Olfert Landt, Marco Kaiser, Richard Molenkamp, Adam Meijer,
  Daniel~K.W. Chu, Tobias Bleicker, Sebastian Br{\"{u}}nink, Julia Schneider,
  Marie~Luisa Schmidt, Daphne~G.J.C. Mulders, Bart~L. Haagmans, Bas {Van Der
  Veer}, Sharon {Van Den Brink}, Lisa Wijsman, Gabriel Goderski, Jean~Louis
  Romette, Joanna Ellis, Maria Zambon, Malik Peiris, Herman Goossens, Chantal
  Reusken, Marion~P.G. Koopmans, and Christian Drosten.
\newblock {Detection of 2019 novel coronavirus (2019-nCoV) by real-time
  RT-PCR}.
\newblock {\em Eurosurveillance}, 25(3):1--8, 2020.

\bibitem{Li2020}
Yan Li and Liming Xia.
\newblock {Coronavirus disease 2019 (COVID-19): Role of chest CT in diagnosis
  and management}.
\newblock {\em American Journal of Roentgenology}, 214(6):1280--1286, jun 2020.

\bibitem{Kong2020}
Weifang Kong and Prachi~P. Agarwal.
\newblock {Chest Imaging Appearance of COVID-19 Infection}.
\newblock {\em Radiology: Cardiothoracic Imaging}, 2(1):e200028, jan 2020.

\bibitem{Fang2020}
Yicheng Fang, Huangqi Zhang, Jicheng Xie, Minjie Lin, Lingjun Ying, Peipei
  Pang, and Wenbin Ji.
\newblock {Sensitivity of chest CT for COVID-19: Comparison to RT-PCR}.
\newblock {\em Radiology}, 296(2):E115--E117, aug 2020.

\bibitem{Chung2020}
Michael Chung, Adam Bernheim, Xueyan Mei, Ning Zhang, Mingqian Huang, Xianjun
  Zeng, Jiufa Cui, Wenjian Xu, Yang Yang, Zahi~A. Fayad, Adam Jacobi, Kunwei
  Li, Shaolin Li, and Hong Shan.
\newblock {CT imaging features of 2019 novel coronavirus (2019-NCoV)}.
\newblock {\em Radiology}, 295(1):202--207, feb 2020.

\bibitem{Bai2020a}
Harrison~X. Bai, Ben Hsieh, Zeng Xiong, Kasey Halsey, Ji~Whae Choi, Thi My~Linh
  Tran, Ian Pan, Lin~Bo Shi, Dong~Cui Wang, Ji~Mei, Xiao~Long Jiang, Qiu~Hua
  Zeng, Thomas~K. Egglin, Ping~Feng Hu, Saurabh Agarwal, Fang~Fang Xie, Sha Li,
  Terrance Healey, Michael~K. Atalay, and Wei~Hua Liao.
\newblock {Performance of Radiologists in Differentiating COVID-19 from
  Non-COVID-19 Viral Pneumonia at Chest CT}.
\newblock {\em Radiology}, 296(2):E46--E54, aug 2020.

\bibitem{Wong2020}
Ho~Yuen~Frank Wong, Hiu Yin~Sonia Lam, Ambrose Ho~Tung Fong, Siu~Ting Leung,
  Thomas Wing~Yan Chin, Christine Shing~Yen Lo, Macy Mei~Sze Lui, Jonan
  Chun~Yin Lee, Keith Wan~Hang Chiu, Tom Wai~Hin Chung, Elaine Yuen~Phin Lee,
  Eric Yuk~Fai Wan, Ivan Fan~Ngai Hung, Tina Poy~Wing Lam, Michael~D. Kuo, and
  Ming~Yen Ng.
\newblock {Frequency and Distribution of Chest Radiographic Findings in
  Patients Positive for COVID-19}.
\newblock {\em Radiology}, 296(2):E72--E78, aug 2020.

\bibitem{Huang2020battle}
Zixing Huang, Shuang Zhao, Zhenlin Li, Weixia Chen, Lihong Zhao, Lipeng Deng,
  and Bin Song.
\newblock {The Battle Against Coronavirus Disease 2019 (COVID-19): Emergency
  Management and Infection Control in a Radiology Department}.
\newblock {\em Journal of the American College of Radiology}, 17(6):710--716,
  jun 2020.

\bibitem{Long2020}
Chunqin Long, Huaxiang Xu, Qinglin Shen, Xianghai Zhang, Bing Fan, Chuanhong
  Wang, Bingliang Zeng, Zicong Li, Xiaofen Li, and Honglu Li.
\newblock {Diagnosis of the Coronavirus disease (COVID-19): rRT-PCR or CT?}
\newblock {\em European Journal of Radiology}, 126:108961, may 2020.

\bibitem{Jacobi2020}
Adam Jacobi, Michael Chung, Adam Bernheim, and Corey Eber.
\newblock {Portable chest X-ray in coronavirus disease-19 (COVID-19): A
  pictorial review}.
\newblock {\em Clinical Imaging}, 64:35--42, aug 2020.

\bibitem{Hu2018}
Zilong Hu, Jinshan Tang, Ziming Wang, Kai Zhang, Lin Zhang, and Qingling Sun.
\newblock {Deep learning for image-based cancer detection and diagnosis - A
  survey}.
\newblock {\em Pattern Recognition}, 83:134--149, nov 2018.

\bibitem{Ayesha2021}
Hareem Ayesha, Sajid Iqbal, Mehreen Tariq, Muhammad Abrar, Muhammad Sanaullah,
  Ishaq Abbas, Amjad Rehman, Muhammad Farooq~Khan Niazi, and Shafiq Hussain.
\newblock {Automatic medical image interpretation: State of the art and future
  directions}.
\newblock {\em Pattern Recognition}, 114, jun 2021.

\bibitem{Barata2021}
Catarina Barata, M.~Emre Celebi, and Jorge~S. Marques.
\newblock {Explainable skin lesion diagnosis using taxonomies}.
\newblock {\em Pattern Recognition}, 110, feb 2021.

\bibitem{Yang2019}
Yunyun Yang, Wenjing Jia, and Yunna Yang.
\newblock {Multi-atlas segmentation and correction model with level set
  formulation for 3D brain MR images}.
\newblock {\em Pattern Recognition}, 90:450--463, jun 2019.

\bibitem{Hannun2019}
Awni~Y. Hannun, Pranav Rajpurkar, Masoumeh Haghpanahi, Geoffrey~H. Tison, Codie
  Bourn, Mintu~P. Turakhia, and Andrew~Y. Ng.
\newblock {Cardiologist-level arrhythmia detection and classification in
  ambulatory electrocardiograms using a deep neural network}.
\newblock {\em Nature Medicine}, 25(1):65--69, jan 2019.

\bibitem{Cheng2010}
H.~D. Cheng, Juan Shan, Wen Ju, Yanhui Guo, and Ling Zhang.
\newblock {Automated breast cancer detection and classification using
  ultrasound images: A survey}.
\newblock {\em Pattern Recognition}, 43(1):299--317, jan 2010.

\bibitem{Rajpurkar2017}
Pranav Rajpurkar, Jeremy Irvin, Kaylie Zhu, Brandon Yang, Hershel Mehta, Tony
  Duan, Daisy Ding, Aarti Bagul, Curtis Langlotz, Katie Shpanskaya, Matthew~P.
  Lungren, and Andrew~Y. Ng.
\newblock {CheXNet: Radiologist-Level Pneumonia Detection on Chest X-Rays with
  Deep Learning}.
\newblock {\em CoRR}, abs/1711.05225, 2017.

\bibitem{Jaiswal2016}
S.~{Jaiswal}, M.~F. {Valstar}, A.~{Gillott}, and D.~{Daley}.
\newblock {Automatic Detection of ADHD and ASD from Expressive Behaviour in
  RGBD Data}.
\newblock In {\em 2017 12th IEEE International Conference on Automatic Face
  Gesture Recognition (FG 2017)}, pages 762--769, 2017.

\bibitem{Souza2019}
Johnatan~Carvalho Souza, Jo{\~{a}}o~Ot{\'{a}}vio {Bandeira Diniz},
  Jonnison~Lima Ferreira, Giovanni~Lucca {Fran{\c{c}}a da Silva},
  Arist{\'{o}}fanes {Corr{\^{e}}a Silva}, and Anselmo~Cardoso de~Paiva.
\newblock {An automatic method for lung segmentation and reconstruction in
  chest X-ray using deep neural networks}.
\newblock {\em Computer Methods and Programs in Biomedicine}, 177:285--296, aug
  2019.

\bibitem{Jeelani2018}
Haris Jeelani, Jonathan Martin, Francis Vasquez, Michael Salerno, and Daniel~S.
  Weller.
\newblock {Image quality affects deep learning reconstruction of MRI}.
\newblock In {\em Proceedings - International Symposium on Biomedical Imaging},
  volume 2018-April, pages 357--360. IEEE Computer Society, may 2018.

\bibitem{Chrysostomou2020}
Charalambos Chrysostomou, Loizos Koutsantonis, Christos Lemesios, and Costas~N.
  Papanicolas.
\newblock {SPECT Imaging Reconstruction Method Based on Deep Convolutional
  Neural Network}.
\newblock {\em 2019 IEEE Nuclear Science Symposium and Medical Imaging
  Conference, NSS/MIC 2019}, oct 2020.

\bibitem{Rudin2018}
Cynthia Rudin.
\newblock {Stop explaining black box machine learning models for high stakes
  decisions and use interpretable models instead}.
\newblock {\em NIPS 2018}, pages 1--15, nov 2018.

\bibitem{Liberati2009}
Alessandro Liberati, Douglas~G. Altman, Jennifer Tetzlaff, Cynthia Mulrow,
  Peter~C. G{\o}tzsche, John~P.A. Ioannidis, Mike Clarke, P.~J. Devereaux, Jos
  Kleijnen, and David Moher.
\newblock {The PRISMA statement for reporting systematic reviews and
  meta-analyses of studies that evaluate health care interventions: Explanation
  and elaboration}.
\newblock {\em PLoS Medicine}, 6(7), 2009.

\bibitem{b8}
Luca Brunese, Francesco Mercaldo, Alfonso Reginelli, and Antonella Santone.
\newblock {Explainable Deep Learning for Pulmonary Disease and Coronavirus
  COVID-19 Detection from X-rays}.
\newblock {\em Computer Methods and Programs in Biomedicine}, 196:105608, 2020.

\bibitem{b13}
Z.~{Han}, B.~{Wei}, Y.~{Hong}, T.~{Li}, J.~{Cong}, X.~{Zhu}, H.~{Wei}, and
  W.~{Zhang}.
\newblock {Accurate Screening of COVID-19 Using Attention-Based Deep 3D
  Multiple Instance Learning}.
\newblock {\em IEEE Transactions on Medical Imaging}, 39(8):2584--2594, 2020.

\bibitem{b16}
M.~R. {Karim}, T.~{Döhmen}, M.~{Cochez}, O.~{Beyan}, D.~{Rebholz-Schuhmann},
  and S.~{Decker}.
\newblock {DeepCOVIDExplainer: Explainable COVID-19 Diagnosis from Chest X-ray
  Images}.
\newblock In {\em IEEE International Conference on Bioinformatics and
  Biomedicine (BIBM)}, pages 1034--1037, 2020.

\bibitem{b18}
Eri Matsuyama.
\newblock {A deep learning interpretable model for novel coronavirus disease
  (COVID-19) screening with chest CT images}.
\newblock {\em Journal of Biomedical Science and Engineering}, 13(7):140--152,
  2020.

\bibitem{b19}
Yujin Oh, Sangjoon Park, and Jong~Chul Ye.
\newblock {Deep Learning COVID-19 Features on CXR using Limited Training Data
  Sets}.
\newblock {\em IEEE Transactions on Medical Imaging}, 0062(c):1--1, 2020.

\bibitem{b20}
Xi~Ouyang, Jiayu Huo, Liming Xia, Fei Shan, Jun Liu, Zhanhao Mo, Fuhua Yan,
  Zhongxiang Ding, Qi~Yang, Bin Song, Feng Shi, Huan Yuan, Ying Wei, Xiaohuan
  Cao, Yaozong Gao, Dijia Wu, Qian Wang, and Dinggang Shen.
\newblock {Dual-Sampling Attention Network for Diagnosis of COVID-19 from
  Community Acquired Pneumonia}.
\newblock {\em IEEE Transactions on Medical Imaging}, 39(XX):1--1, 2020.

\bibitem{b21}
Tulin Ozturk, Muhammed Talo, Eylul~Azra Yildirim, Ulas~Baran Baloglu, Ozal
  Yildirim, and U.~{Rajendra Acharya}.
\newblock {Automated detection of COVID-19 cases using deep neural networks
  with X-ray images}.
\newblock {\em Computers in Biology and Medicine}, 121(April):103792, 2020.

\bibitem{b26}
S.~{Tabik}, A.~{Gómez-Ríos}, J.~L. {Martín-Rodríguez},
  I.~{Sevillano-García}, M.~{Rey-Area}, D.~{Charte}, E.~{Guirado}, J.~L.
  {Suárez}, J.~{Luengo}, M.~A. {Valero-González}, P.~{García-Villanova},
  E.~{Olmedo-Sánchez}, and F.~{Herrera}.
\newblock {COVIDGR Dataset and COVID-SDNet Methodology for Predicting COVID-19
  Based on Chest X-Ray Images}.
\newblock {\em IEEE Journal of Biomedical and Health Informatics},
  24(12):3595--3605, 2020.

\bibitem{b27}
Nikos Tsiknakis, Eleftherios Trivizakis, Evangelia Vassalou, Georgios
  Papadakis, Demetrios Spandidos, Aristidis Tsatsakis, Jose
  S{\'{a}}nchez‑Garc{\'{i}}a, Rafael L{\'{o}}pez‑Gonz{\'{a}}lez, Nikolaos
  Papanikolaou, Apostolos Karantanas, and Kostas Marias.
\newblock {Interpretable artificial intelligence framework for COVID‑19
  screening on chest X‑rays}.
\newblock {\em Experimental and Therapeutic Medicine}, pages 727--735, 2020.

\bibitem{b28}
Ferhat Ucar and Deniz Korkmaz.
\newblock {COVIDiagnosis-Net: Deep Bayes-SqueezeNet based diagnosis of the
  coronavirus disease 2019 (COVID-19) from X-ray images}.
\newblock {\em Medical Hypotheses}, 140(April):109761, 2020.

\bibitem{b29}
Linda Wang, Zhong~Qiu Lin, and Alexander Wong.
\newblock {COVID-Net: a tailored deep convolutional neural network design for
  detection of COVID-19 cases from chest X-ray images}.
\newblock {\em Scientific Reports}, 10(1):19549, Nov 2020.

\bibitem{b30}
Y.~H. Wu, S.~H. Gao, J.~Mei, J.~Xu, D.~P. Fan, R.~G. Zhang, and M.~M. Cheng.
\newblock {{J}{C}{S}: {A}n {E}xplainable {C}{O}{V}{I}{D}-19 {D}iagnosis
  {S}ystem by {J}oint {C}lassification and {S}egmentation}.
\newblock {\em IEEE Trans Image Process}, 30:3113--3126, 2021.

\bibitem{b5}
Md~Manjurul Ahsan, Kishor~Datta Gupta, Mohammad~Maminur Islam, Sajib Sen,
  Md.~Lutfar Rahman, and Mohammad~Shakhawat Hossain.
\newblock {Study of Different Deep Learning Approach with Explainable AI for
  Screening Patients with COVID-19 Symptoms: Using CT Scan and Chest X-ray
  Image Dataset}.
\newblock {\em arXiv}, page 2007.12525, 2020.

\bibitem{b6}
Nikita Albert.
\newblock {Evaluation of Contemporary Convolutional Neural Network
  Architectures for Detecting COVID-19 from Chest Radiographs}.
\newblock {\em arXiv}, page 2007.01108, 2020.

\bibitem{b7}
Plamen~P Angelov and Eduardo~A Soares.
\newblock {Explainable-By-Design Approach For Covid-19 Classification Via
  CT-Scan}.
\newblock {\em medRxiv}, (July):2020.04.24.20078584, 2020.

\bibitem{b9}
Soumick Chatterjee, Fatima Saad, Chompunuch Sarasaen, Suhita Ghosh, Rupali
  Khatun, Petia Radeva, Georg Rose, Sebastian Stober, Oliver Speck, and Andreas
  N{\"{u}}rnberger.
\newblock {Exploration of Interpretability Techniques for Deep COVID-19
  Classification using Chest X-ray Images}.
\newblock {\em arXiv}, page 2006.02570, 2020.

\bibitem{b10}
Joseph~Paul Cohen, Lan Dao, Paul Morrison, Karsten Roth, Yoshua Bengio, Beiyi
  Shen, Almas Abbasi, Mahsa Hoshmand-Kochi, Marzyeh Ghassemi, Haifang Li, and
  Tim~Q Duong.
\newblock {Predicting COVID-19 Pneumonia Severity on Chest X-ray with Deep
  Learning}.
\newblock {\em arXiv}, 8(December 2019):2005.11856, 2020.

\bibitem{b11}
Biraja Ghoshal and Allan Tucker.
\newblock {Estimating Uncertainty and Interpretability in Deep Learning for
  Coronavirus (COVID-19) Detection}.
\newblock {\em arXiv}, page 2003.10769, 2020.

\bibitem{b12}
Ophir Gozes, Ma~Frid, Hayit Greenspan, and D~Patrick.
\newblock {Rapid AI Development Cycle for the Coronavirus ( COVID-19 ) Pandemic
  : Initial Results for Automated Detection {\&} Patient Monitoring using Deep
  Learning CT Image Analysis}.
\newblock {\em arXiv}, page 2003.05037, 2020.

\bibitem{b15}
Amit~Kumar Jaiswal, Prayag Tiwari, Vipin~Kumar Rathi, Jia Qian, Hari~Mohan
  Pandey, and Victor Hugo~C Albuquerque.
\newblock {COVIDPEN: A Novel COVID-19 Detection Model using Chest X-Rays and CT
  Scans}.
\newblock {\em medRxiv}, page 2020.07.08.20149161, 2020.

\bibitem{b17}
Shahin Khobahi, Chirag Agarwal, and Mojtaba Soltanalian.
\newblock {CoroNet: A Deep Network Architecture for Semi-Supervised Task-Based
  Identification of COVID-19 from Chest X-ray Images}.
\newblock {\em medRxiv}, page 2020.04.14.20065722, 2020.

\bibitem{b23}
Laboni Sarker, Mohaiminul Islam, Tanveer Hannan, Ahmed Zakaria, Zakaria Ahmed,
  and Ahmed Zakaria.
\newblock {COVID-DenseNet: A deep learning architecture to detect COVID-19 from
  chest radiology images}.
\newblock {\em Preprints}, (May):2020050151, 2020.

\bibitem{b24}
Vishal Sharma and Curtis Dyreson.
\newblock {COVID-19 detection using Residual Attention Network an Artificial
  Intelligence approach}.
\newblock {\em arXiv}, page 2006.16106, 2020.

\bibitem{b25}
Alberto Signoroni, Mattia Savardi, Sergio Benini, Nicola Adami, Riccardo
  Leonardi, Paolo Gibellini, Filippo Vaccher, Marco Ravanelli, Andrea Borghesi,
  Roberto Maroldi, and Davide Farina.
\newblock {End-to-end learning for semiquantitative rating of COVID-19 severity
  on Chest X-rays}.
\newblock {\em arXiv}, page 2006.04603, 2020.

\bibitem{b31}
Maryam Zokaeinikoo, Prasenjit Mitra, Soundar Kumara, and Pooyan Kazemian.
\newblock {AIDCOV: An Interpretable Artificial Intelligence Model for Detection
  of COVID-19 from Chest Radiography Images}.
\newblock {\em medRxiv}, page 2020.05.24.20111922, 2020.

\bibitem{Yoon2020}
Soon~Ho Yoon, Kyung~Hee Lee, Jin~Yong Kim, Young~Kyung Lee, Hongseok Ko,
  Ki~Hwan Kim, Chang~Min Park, and Yun~Hyeon Kim.
\newblock {Chest radiographic and ct findings of the 2019 novel coronavirus
  disease (Covid-19): Analysis of nine patients treated in Korea}.
\newblock {\em Korean Journal of Radiology}, 21(4):498--504, apr 2020.

\bibitem{Lee2020}
Edward~Y. Lee.
\newblock {Pediatric Interstitial (Diffuse) Lung Disease}.
\newblock In {\em Imaging in Pediatric Pulmonology}, pages 145--197. Springer
  International Publishing, 2020.

\bibitem{Wynants2020}
Laure Wynants, Ben {Van Calster}, Marc~M.J. Bonten, Gary~S. Collins,
  Thomas~P.A. Debray, Maarten {De Vos}, Maria~C. Haller, Georg Heinze,
  Karel~G.M. Moons, Richard~D. Riley, Ewoud Schuit, Luc~J.M. Smits, Kym~I.E.
  Snell, Ewout~W. Steyerberg, Christine Wallisch, and Maarten {Van Smeden}.
\newblock {Prediction models for diagnosis and prognosis of covid-19 infection:
  Systematic review and critical appraisal}.
\newblock {\em The BMJ}, 369, 2020.

\bibitem{Cheplygina2018}
Veronika Cheplygina.
\newblock {Cats or CAT scans: transfer learning from natural or medical image
  source datasets?}
\newblock {\em Current Opinion in Biomedical Engineering}, 9:21--27, oct 2018.

\bibitem{Kim2017}
Hak~Gu Kim, Yeoreum Choi, and Yong~Man Ro.
\newblock {Modality-bridge Transfer Learning for Medical Image Classification}.
\newblock {\em Proceedings - 2017 10th International Congress on Image and
  Signal Processing, BioMedical Engineering and Informatics, CISP-BMEI 2017},
  2018-January:1--5, aug 2017.

\bibitem{Albahri2020}
O.S. Albahri, A.A. Zaidan, A.S. Albahri, B.B. Zaidan, Karrar~Hameed
  Abdulkareem, Z.T. Al-qaysi, A.H. Alamoodi, A.M. Aleesa, M.A. Chyad, R.M.
  Alesa, L.C. Kem, Muhammad~Modi Lakulu, A.B. Ibrahim, and Nazre~Abdul Rashid.
\newblock Systematic review of artificial intelligence techniques in the
  detection and classification of covid-19 medical images in terms of
  evaluation and benchmarking: Taxonomy analysis, challenges, future solutions
  and methodological aspects.
\newblock {\em Journal of Infection and Public Health}, 13(10):1381--1396,
  2020.

\bibitem{biecekburzykowski}
Przemyslaw Biecek and Tomasz Burzykowski.
\newblock {\em {Explanatory Model Analysis}}.
\newblock Chapman and Hall/CRC, New York, 2021.

\bibitem{Holzinger2017}
Andreas Holzinger, Chris Biemann, Constantinos~S. Pattichis, and Douglas~B.
  Kell.
\newblock {What do we need to build explainable {AI} systems for the medical
  domain?}
\newblock {\em CoRR}, abs/1712.09923:1712.09923, 2017.

\bibitem{Samek2017}
Wojciech Samek, Thomas Wiegand, and Klaus{-}Robert M{\"{u}}ller.
\newblock {Explainable Artificial Intelligence: Understanding, Visualizing and
  Interpreting Deep Learning Models}.
\newblock {\em CoRR}, abs/1708.08296:1708.08296, 2017.

\bibitem{DeYoung2019}
Jay DeYoung, Sarthak Jain, Nazneen~Fatema Rajani, Eric Lehman, Caiming Xiong,
  Richard Socher, and Byron~C. Wallace.
\newblock {ERASER}: {A} benchmark to evaluate rationalized {NLP} models.
\newblock In {\em Proceedings of the 58th Annual Meeting of the Association for
  Computational Linguistics}, pages 4443--4458, Online, jul 2020. Association
  for Computational Linguistics.

\bibitem{gawande2011checklist}
Atul Gawande.
\newblock {\em {The Checklist Manifesto: How to Get Things Right}}.
\newblock Profile, 2011.

\end{thebibliography}

\end{document}